\documentclass[12pt, a4paper]{article}
\usepackage[english]{babel} 
\usepackage{amsmath,amsfonts,amsthm} 
\usepackage{graphicx}
\usepackage{hyperref}
\usepackage{xcolor}
\usepackage{subcaption}
\usepackage{lscape}
\usepackage{natbib}
\usepackage{longtable}
\usepackage{fancyhdr} 

\numberwithin{equation}{section} 
\numberwithin{figure}{section} 
\numberwithin{table}{section} 

\usepackage[margin=1in]{geometry}
\fancypagestyle{firststyle}
{
   \fancyhf{}
   \newcommand{\changefont}{
    \fontsize{9}{11}\selectfont
}
   \fancyhead[L]{\changefont This is a post-peer-review, pre-copyedit version of an article published in Energy Economics. The final authenticated version is available online at: \url{http://dx.doi.org/10.1016/j.eneco.2019.06.021}}
}

\title{\normalfont  \Large Regularization Approach for Network Modeling of German Power Derivative Market
\thanks{Financial support from the IRTG 1792 "High Dimensional Non Stationary Time Series", as well as the Czech Science Foundation under grant no. 19-28231X, the Yushan Scholar Program and the European Union's Horizon 2020 research and innovation program "FIN-TECH: A Financial supervision and Technology compliance training programme" under the grant agreement No 825215 (Topic: ICT-35-2018, Type of action: CSA), Humboldt-Universit\"at zu Berlin, is gratefully acknowledged.}}
\author{\normalsize Shi Chen \thanks{Corresponding author, Chair of Statistics and Econometrics, Karlsruher Institut f\"ur Technologie, Bl\"ucherstr.17, $76185$ Karlsruhe, Germany. Email: shi.chen@kit.edu}, 
Wolfgang Karl H{\"a}rdle\thanks{School of Business and Economics, Humboldt-Universit\"at zu Berlin, Unter den Linden 6, 10099 Berlin, Germany; Singapore Management University, 50 Stamford Road, Singapore 178899}, 
Brenda L\'opez Cabrera \thanks{School of Business and Economics, Humboldt-Universit\"at zu Berlin, Unter den Linden 6, 10099 Berlin, Germany}
}

\date{}

\begin{document}
\maketitle

\thispagestyle{firststyle}

\begin{abstract}

In this paper we propose a regularization approach for network modeling of German power derivative market. To deal with the large portfolio, we combine high-dimensional variable selection techniques with dynamic network analysis. The estimated sparse interconnectedness of the full German power derivative market, clearly identify the significant channels of relevant potential risk spillovers. Our empirical findings show the importance of interdependence between different contract types, and identify the main risk contributors. We further observe strong pairwise interconnections between the neighboring contracts especially for the spot contracts trading in the peak hours, its implications for regulators and investors are also discussed. The network analysis of the full German power derivative market helps us to complement a full picture of system risk, and have a better understanding of the German power market functioning and environment.

\end{abstract}
\normalsize{\textbf{JEL}: C1, Q41, Q47

\newpage
\section{Introduction}
Affordable and reliable energy supply is essential for industrial growth. Achieving these in times of growing demand, raw materials shortage and climate change poses challenges. Germany's power system for the industry and the consumers is undergoing radical change, this transformation is being driven by the restructuring of electricity supply and by intense competition between suppliers (see \cite{BMWi}, \cite{spiecker2014impact}, \cite{seifert2016technical}, \cite{grossi2017impact}, \cite{sinn2017buffering} and among others). However, the ongoing expansion of renewable energy and the phase-out of nuclear energy for power generation will change the composition of the electricity mix, which in return, will generate pricing signals affecting the electricity trading (e.g. \cite{benhmad2018photovoltaic}, \cite{ketterer2014impact}, \cite{ballester2015effects}, \cite{paraschiv2014impact}). As we know, electricity is the commodity that should be supplied immediately. Unlike coal, oil, gas or other typical commodities, electricity cannot be stored. This results in the price of electricity being volatile and very depended on a secure supply. To hedge against the uncertainty arisen in the market, we study the system-wide market risk of the whole German power derivative market. Therefore energy companies may invest in both electricity spot and derivative markets to diversity their existing portfolios. As electricity grids worldwide also begin relying more heavily on renewable energy sources, analysis of German power market thus provides useful insights for power generation companies and transmission organizations across the globe. 

 However, the number of variables and relevant factors is typically huge. A properly designed subset selection has to be employed to identify the most informative power contracts to representing energy market risk. The German power market is highly interconnected with a dense and wide range of electricity contracts, this motivates us to build up an ultra high dimensional network and investigate its sparse property. To better understand the interaction between power contracts, the iterated sure independence screening (iterated-SIS, \cite{fan2009ultrahigh}) method combined with regularization estimators are applied to estimate the sparse web of connections. Our network of interest is constructed in the context of time series based on vector autoregression (VAR) models, the iterated-SIS method is important when building VAR models since the number of parameters to estimate increases quadratically in the number of variables included. To quantify the associations between individual power contract and energy exchange market, the network we constructed is obtained from the forecast error variance decomposition  (FEVD) based on VAR estimates in the framework of \cite{koop1996impulse} and \cite{pesaran1998generalized}. This kind of connectedness measure is also used by \cite{diebold2009measuring}, \cite{diebold2014network} for conceptualizing and empirically measuring weighted, directed networks at a variety of levels. They proposed this variance decomposition networks as tools for quantifying and ranking the systemic risk of individual component in a portfolio.

In this paper we investigate the concept of connectedness in a realistic high dimensional framework, which is important for system-wide risk measurement and management. We aim to obtain a sparse network in which nodes represent power contracts and links represent the magnitude of connectedness, local shocks and events can therefore be easily amplified and turned into global events. 
While estimates of the network yield the qualitative links between power contracts, individual impact from specific contract can be estimated and speculated accordingly. Hence the risk contribution from the market component can be identified, this will help us to learn more about the German power market functioning and environment. Following \cite{diebold2014network}, \cite{demirer2018estimating}, \cite{hautsch2014financial} and similar connectedness literature, the risk refers to the uncertainty arisen in the system, and it measures the amount of future uncertainty caused by other component in the system. For example, the market uncertainty may be caused by such as economic and financial uncertainties, or weather conditions, energy prices and regulation. While the systemic risk network yields qualitative information on risk channels and roles of assets within the constructed portfolio, the risk is quantified by its contribution to the forecast error variance of other components in the whole system. Therefore we are able to capture the systemic risk of the component by summing up its total contribution.
Understanding the risk transmission channels for investors is of great importance, for example, our results show that day-ahead spot power contracts that bidding between 9am and 13am are in the core of the German energy power market, the key derivatives in connecting markets can be identified.

The rest of the paper proceeds as follows. Section \ref{sec: energymarket} reviews the relevant literature and introduces the German energy market. In section \ref{sec: theomodel}, we describe in details how the regularization approach is applied to estimate the large portfolio and how the network is constructed. Section \ref{sec: empiricalres} reports the data and discusses the model selection result. Section \ref{sec: networkana} presents a static analysis of full-sample connectedness. Section \ref{sec: dynamicnet} provides the empirical results of the dynamic network. Finally section \ref{sec: conc} concludes.
\section{Background}\label{sec: energymarket}
\subsection{Related literature}
As a tradable commodity electricity is relatively new, its dynamic properties have been analyzed with many different approaches, some recent contributions are \cite{weron2007modeling}, \cite{geman2006understanding}, \cite{bierbrauer2007spot} and \cite{knittel2005empirical}. There is also a strand of literature that analyzes the multivariate behavior of electricity prices. For examples, \cite{higgs2009modelling} examines the inter-relationships of wholesale spot electricity prices across four Australasian markets by a multivariate GARCH model. \cite{henriques2008oil} develops a four variable VAR model to explain the dependence structure of a variety of energy equities, where they find that shocks to technology actually have a larger impact on the stock prices of alternative energy companies than do oil prices. \cite{castagneto2014dynamic} studies  the interactions of a representative sample of 13 European (EU) electricity spot prices with dynamic Granger-causal networks.

We take our starting point in the energy literature based on a vector regression framework, where the network of interest is constructed based on \cite{diebold2009measuring}, \cite{diebold2014network}. Most relevant studies explore the relationship between oil and energy equity prices in terms of volatility spillovers, they estimate the implied volatility linkages across markets as source of future uncertainty. For example \cite{du2011speculation} conducts a Bayesian analysis to explain volatility spillovers among crude oil and various economic factors. \cite{arouri2012impacts} investigates the volatility spillovers between oil and stock markets in Europe using VARDGARCH approach. \cite{sadorsky2012correlations} applies a multivariate GARCH model to estimate the volatility spillovers between oil prices and the stock prices of clean energy/technology companies. \cite{joo2017oil} examines the time-varying causal relationship between the stock and crude oil price uncertainties using a DCC GARCH-in-Mean specification. More empirical work are \cite{diebold2012better}, \cite{reboredo2014volatility}, \cite{maghyereh2016directional}, \cite{awartani2016connectedness}, \cite{zhang2017oil}, \cite{apergis2017good}, and among others. There is also a fairly sizable literature exploring the electricity market integration using vector regression framework, for example \cite{worthington2005transmission}, \cite{zachmann2008electricity}, \cite{bunn2010integration}, \cite{balaguer2011cross}, \cite{bockers2014extent}, \cite{castagneto2014dynamic}.

However, relatively little research has focused on the systemic directional interaction between energy equities. A recent study by \cite{lundgren2018connectedness} is the first to analyze the connectedness network among different energy asset classes, their analysis examines the connectedness network among renewable energy stock, four investments, and uncertainties. \cite{demirer2018estimating} uses Lasso method to select, shrink and estimate a high-dimensional network. Other empirical work are with more focus on financial banking contexts, like \cite{yi2018volatility} uses the VARX-L framework developed by \cite{nicholson2017varx} to conduct static and dynamic volatility spillovers among cryptocurrencies. More relevant work are \cite{wang2018volatility}, \cite{acharya2012capital}, \cite{hautsch2014financial}, \cite{giglio2016systemic}, \cite{babus2016formation}, \cite{brownlees2016srisk}, \cite{acharya2017measuring} and among others.

However, almost all existing energy literature are based only on moderate dimensions. This motivates us to examine the systemic risk transmission channels in a realistic high dimensional framework. The main argument is the contracts trading in both spot and derivative markets share the same underlying contracts, and it is therefore natural to consider a high dimensional portfolio. High-dimensional statistical problems arise from diverse fields of scientific research and technological development, including energy markets. The traditional idea of best subset selection methods is computationally too expensive for many modern statistical applications. Variable selection techniques have therefore been successfully developed in recent years and they indeed play a pivotal role in contemporary statistical learning and techniques. Researchers have proposed various regularized estimators with different penalty terms, a preeminent example being the least absolute selection and shrinkage operator (Lasso) of \cite{tibshirani1996}. In recent years, Lasso has been extended to high-dimensional case, see \cite{lasso2009}. Other popular methods contribute to the literature,  such as smoothly clipped absolute deviation (SCAD) \cite{fan2001variable}, adaptive Lasso of \cite{zou2006adaptive}, elastic net estimator of \cite{zou2005regularization}, Dantzig selector of \cite{candes2007dantzig}. In an ultra high-dimensional case where the dimensionality of the model is allowed to grow exponentially in the sample size, it is helpful to begin with screening to delete some significantly irrelevant variables from the model. \cite{fan2008sure} introduce a method called sure independence screening for this goal. Even when the regularity conditions may not hold, \cite{fan2009ultrahigh} extend the iterated-SIS method to work by iteratively performing feature selection to recruit a small number of features. Furthermore, the asymptotic properties of Lasso for high-dimensional time series have been considered by \cite{loh2011high} and \cite{wu2016performance}. \cite{kock_var2} establishes the high-dimensional VAR estimation with focus on Lasso and adaptive Lasso. \cite{basu2015regularized} investigates the theoretical properties of regularized estimates in sparse high-dimensional time series models when the data are generated from a multivariate stationary Gaussian process.

We address the above high dimensional problem by combining a regularization approach with classic VAR model. In doing so, we contribute to the energy literature in several ways: First, we extend the current literature to investigate the sparse linkage between energy equities in a very large portfolio. Second, we identify the main risk contributor to help investor diversity their existing portfolio rather than having large holdings of individual electricity contract. Third, our study may further be extended by including renewable energy assets and oil price across European energy market, this may provide a better understanding of the overall energy market. In addition, our study can also be applied to investigate electricity market integration by estimating the total connectedness for a wide range of components.

\subsection{Overview of German power market}\label{subsec: market}
\paragraph{German electricity market}
The German electricity market is Europe's largest, with annual power consumption of around 530 TWh and a generation capacity of 184 GW. As a net energy exporter, the export capacity of Germany is expected to continue to grow as planned interconnections expand cross-border transmission capacity with several neighboring countries. Germany has significant interconnection capacity with neighboring EU member states as well. It is interconnected with Austria, Switzerland, the Czech Republic, Denmark, France, Luxembourg, the Netherlands, Poland, and Sweden. To maintain stable and reliable supply of electricity, the so-called Transmission system operators (TSOs) keep control power available. Primary control, secondary control, and tertiary control reserve are procured by the respective TSOs within a non-discriminatory control power market in accordance with the requirements of the Federal Cartel Office. Demand for control energy is created when the sum of power generated varies from the actual load caused by unforeseeable weather fluctuations in the case of renewable energies.

Electricity is traded on the exchange and over the counter. Standardized products are bought and sold in a transparent process on the exchange, which, for Germany, is the European Energy Exchange EEX in Leipzig, the European Energy Exchange EPEX SPOT in Paris and the Energy Exchange Austria (EXAA) in Vienna. The European Energy Exchange (EEX) is the leading energy exchange in Europe. It develops, operates and connects secure, liquid and transparent markets for energy and commodity products. Contracts on power, coal and CO2 emission allowances as well as freight and agricultural products are traded or registered for clearing on EEX. EPEX SPOT, Powernext, Cleartrade Exchange (CLTX) and Gaspoint Nordic are also members of EEX Group. The German wholesale electricity market is broadly made up of three elements, a forward market, a day-ahead market and an intra-day market. These submarkets generate the pricing signal which electricity production and consumption align to. The objective of this paper is to analyse the interaction of different future contracts traded in the forward market, whether forward market is influenced by market power of spot prices traded in EPEX market.

\begin{figure}[h]
\centering
\includegraphics[scale=0.20]{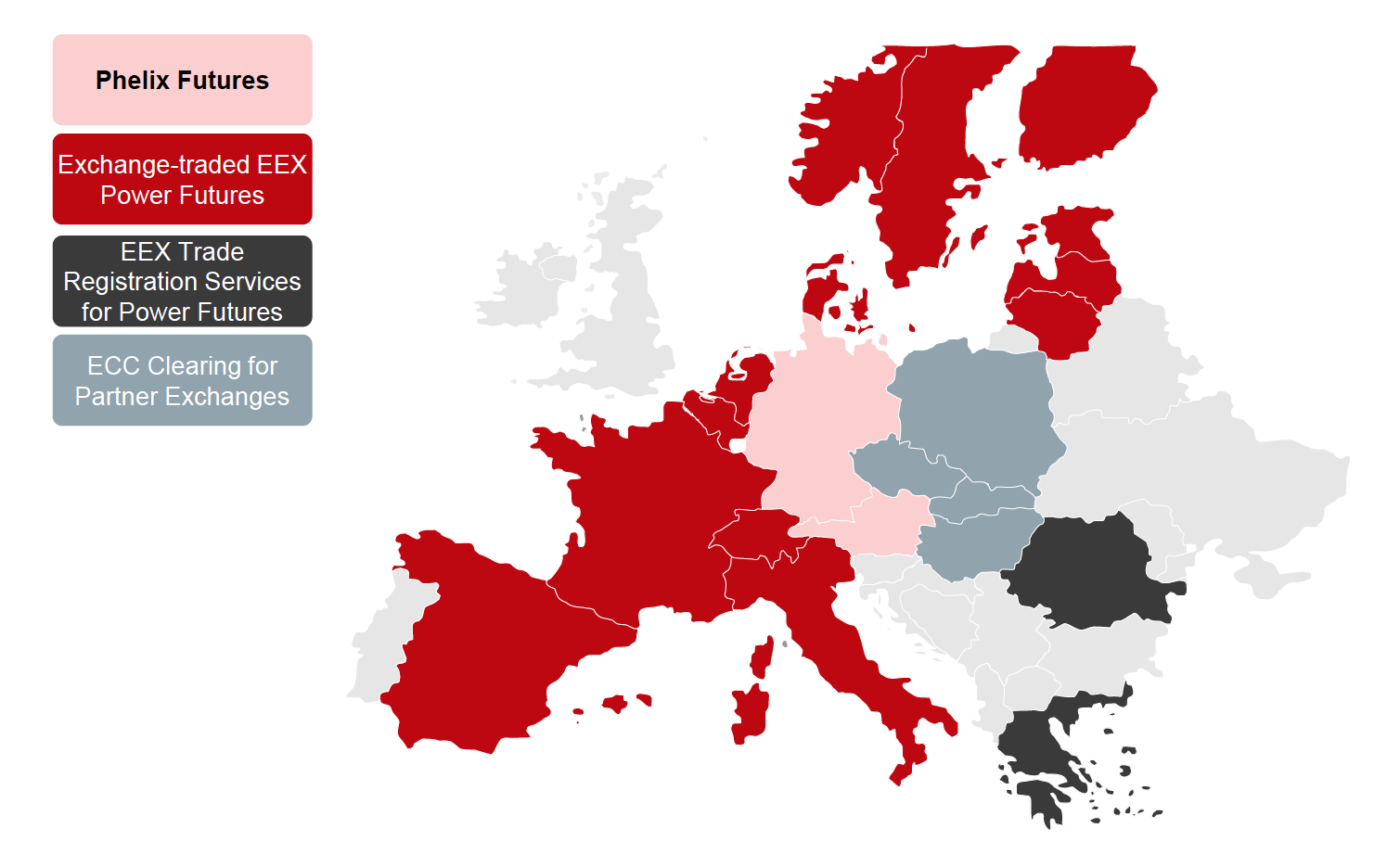}
\caption{The distribution of European power derivatives in EEX market. Source: EEX website}\label{fig: derimarket}
\end{figure}

Electricity providers and electricity purchasers submit their bids in their national day-ahead market zones. The exchange price on the day-ahead market is determined jointly for coupled markets. Electricity providers and electricity purchasers submit their bids in their national day-ahead market zones. In an iterative process, the demand for electricity in the market zone is served by the lowest price offers of electricity from all the market areas until the capacity of the connections between the market zones (cross-border inter-connectors) is fully utilized. As long as the cross-border inter-connectors have sufficient capacity, this process aligns the prices in the coupled market areas. On account of market coupling, the national power demand is covered by the international offers with lowest prices. The upshot is that on the whole less capacity is required to meet the demand. As shown in Figure \ref{fig: derimarket}, Phelix Future, as the product traded in Germany, is a financial derivatives contract settling against the average power spot market prices of future delivery periods for the German/Austrian market area.

\paragraph{Phelix futures}  
Electricity supply deliveries in the forward market can be negotiated up to seven years in advance, but for liquidity reasons typically only look out three years, and in fact one year ahead futures are traded at most. The Phelix Future is a financial derivatives contract referring to the average power spot market prices of future delivery periods of the German/Austrian market area.

As the most liquid contract and benchmark for European power trading, the underlying of these future contracts is the Physical Electricity Index determined daily by EPEX Spot Exchange for base and peak load profiles. To be more specific, the Phelix Base contract is average price of the hours 1 to 24 for electricity traded on spot market, while the Phelix Peak is the average price of the hours 9 to 20 for electricity traded on spot market. EEX offers continuous trading and trade registration of financially fulfilled Phelix Futures, with Day/Weekend Futures, Week Futures, Month Futures, Quarter Futures and Year Futures available. 

The time series of Phelix day base and Phelix day peak prices are displayed in Figure \ref{fig:plot_prices}.  Phelix day peak exhibit a larger volatility and more
pronounced spikes than the Phelix day base. This is not surprising, since the Phelix day peak corresponds to hours with high and variable demand. Both price series exhibit positive skewness and an excess kurtosis of about 1, implying a heavy-tailed unconditional distribution that is skewed to the right. 

\begin{figure}[h]
\centering
\includegraphics[width=0.65\textwidth]{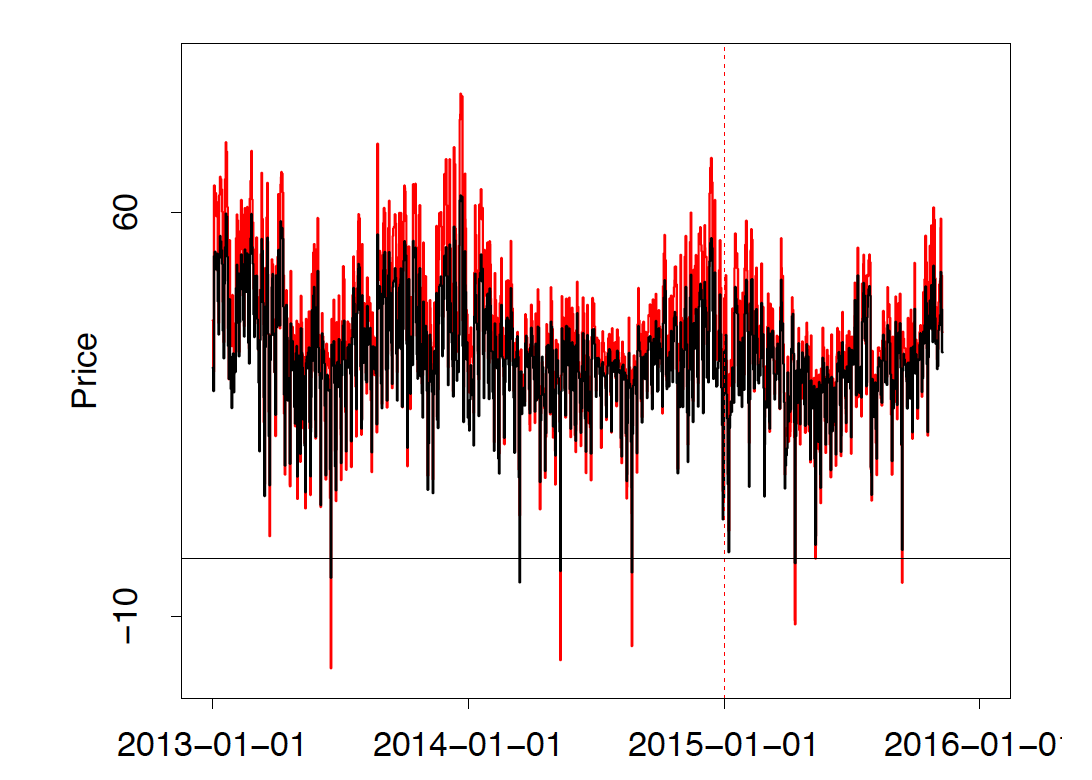}
\caption{Phelix day base (black) and Phelix day peak (red) index from 2013-01-01 to 2015-10-31. The red dotted line marks the end of the in-sample period.}\label{fig:plot_prices}
\end{figure}

In addition, the Phelix market is also successfully connected to other European power markets. The products of Location Spread enables members to trade price differences between markets, thus enabling participants to benefit from improved liquidity and tighter spreads, for instance, Phelix / French Power, Italian / Phelix Power, Phelix / Nordic Power and Phelix / Swiss Power. For the empirical work of this paper, we use the Phelix Future data to find price drivers and important variables in the big system we construct. The decision-making mechanism of energy companies will also be explored. 

\section{Econometric Model}\label{sec: theomodel}
\subsection{Basic model}

When there is a high-dimensional portfolio consisting of various power derivative contracts, standard methods are intractable and therefore require novel statistical methods. Here we are interested in addressing the following research questions, how all these contracts interact with each other? Which variables are crucial for the whole system? However, due to the large number of variables in the system, some sparsity assumption must be imposed for the sake of an accurate estimate. The large dimensionality in our model comes from not only the varieties of power derivative products, but also the large lag order in VAR model to avoid the correlation of error terms.

The standard VAR($p$) model with lag order $p$ is constructed according to \cite{lutkepohl2005new},
\begin{eqnarray}
  y_{t} &=& \nu + A_{1}y_{t-1}+ A_{2}y_{t-2}+\dots+ A_{p}y_{t-p}+u_{t} \nonumber\\
  &=& \nu + \left( A_{1}, A_{2}, \ldots, A_{p} \right) \left( y_{t-1}^{\top}, y_{t-2}^{\top},\ldots, y_{t-p}^{\top} \right)^{\top}+u_{t}
\label{equ: var1}
\end{eqnarray}
where $y_{t} = ( y_{1t}, y_{2t}, \ldots, y_{Kt})^{\top}$ is a $(K\times1 )$ random vector consisting $K$ variables at time $t$, $t=1,\ldots, T$. $A_{i}$ are unknown ($K \times K$) coefficient matrices. $\nu$ is a $(K\times1)$ vector of intercept terms, $u_{t} = ( u_{1t}, u_{2t}, \ldots, u_{Kt} )^{\top}$ is a $K$-dimensional innovation process.
Define
\begin{eqnarray}
Y &=& (y_{1}, y_{2}, \ldots, y_{T}) \nonumber \\
B &=& (\nu, A_{1}, A_{2}, \ldots, A_{p}) \nonumber \\
Z_{t} &=& \left( 1, y_{t}, y_{t-p+1} \right)^{\top} \nonumber \\
Z &=& (Z_{0}, Z_{1}, \ldots,Z_{T-1})
\end{eqnarray}
For multivariate case, rewrite equation \eqref{equ: var1} as
\begin{eqnarray}
  Y=BZ + U
\label{var2}
\end{eqnarray}
where $U=(u_{1}, u_{2}, \ldots ,u_{T})$. The compact form of \eqref{var2} is
\begin{eqnarray}
  \label{ls1}
  vec(Y)&=&(Z^{\top} \otimes I_{K})vec(B)+vec(U) \nonumber \\ 
 \textbf{y} &=& (Z^{\top} \otimes I_{K}) \mathcal{\beta} + \textbf{u} = \textbf{X} \mathcal{\beta} + \textbf{u}
\end{eqnarray}

The dimension of model \eqref{ls1} to be estimated is $pK^2$ and the number of observations is $KT$. The ratio $\frac{Kp}{T}$ could be large due to the reasons mentioned earlier, which deteriorates the accuracy of final estimate. Worse still, if $Kp>T$, the model becomes high-dimensional with more unknown parameters than observations. Therefore we require teniques other than traditional OLS method. That's why most existing energy literature are based only on moderate dimensions.

Here we use variable selection technique, for example Lasso, to estimate the model. Besides, under normal assumption of error term, the upper bound of error in estimation is positively correlated in $\frac{\log(K^{2}p)}{T}$, part of oracle inequality. the estimation results can be further developed by adding one more step of sure independent screening (SIS) before variable selection step. Another advantage of SIS is that it could mitigate the problem caused by multicollinearity, which is common in time series setting. The techniques introduced in the proceeding paragraph are of great importance in the sense that the true underlying model has a sparse representation.

\subsection{Regularization estimator and iterated-SIS algorithm}\label{subsec: reis}
\paragraph{Regularization approach}
Variable selection is an important tool for the linear regression analysis. A popular method is the Lasso estimator of \citet{tibshirani1996}, which can be viewed to simultaneously perform model selection and parameter estimation. Related literature includes bridge regression studied by \citet{frank1993statistical} and \citet{fu1998penalized}, the least angle regression of \citet{efron2004least} and adaptive Lasso proposed by \citet{zou2006adaptive}. Another remarkable example is a smoothly clipped absolute deviation (SCAD) penalty for variable selection proposed by \citet{fan2001variable}, they proved its oracle properties. 

Let us start with consider model estimation and variable selection for equation \eqref{ls1},
\begin{eqnarray} 
 \textbf{y} = \textbf{X} \mathcal{\beta} + \textbf{u}
\end{eqnarray}

The least square estimate is obtained via minimizing $\| y - X\beta \|^{2}$, where the ordinary least squares (OLS) gives nonzero estimates $  \omega = X^{\top}y$ to all coefficients. Normally best subset selection is implemented to select significant variables, but the traditional idea of best subset selection methods is computationally too expensive for many statistical applications. Therefore the penalized least square  with a penalty term that is separable with respect to the estimated parameter $\hat{\beta}$ is considered here. In this paper we consider two popular estimators, Lasso and SCAD.

The Lasso is a regularization technique for simultaneous estimation and variable selection, with the estimator given by,
\begin{eqnarray}
\hat{\beta}_{LASSO} &=& \operatorname{arg}\,\underset{\beta}{\operatorname{min}} \Vert y - X\beta \Vert ^{2} + \lambda \sum_{j=1}^{p} \vert \beta_{j} \vert \nonumber\\ 
                 &=&  \operatorname{arg}\,\underset{\beta}{\operatorname{min}} \Vert y - \sum_{j=1}^{p} x_{j}\beta_{j} \Vert ^{2} + \lambda \sum_{j=1}^{p} \vert \beta_{j} \vert 
                 \label{equ: lasso}
\end{eqnarray}
where $\lambda$ is the tuning parameter. The second term in equation \eqref{equ: lasso} is known as the $\ell_{1}$-penalty. The idea behind Lasso is the coefficients shrink toward 0 as $\lambda$ increases. When $\lambda$ is sufficiently large, some of the estimated coefficients are exactly zero. The estimation accuracy comes from the trade-off between estimation variance and the bias. Lasso is the penalized least square estimates with the $\ell_{1}$ penalty in the general least squares and likelihood settings. Furthermore, the $\ell_{2}$ penalty results in a ridge regression and $\ell_{p}$ penalty will lead to a bridge regression. 


We proceed to the SCAD method. In the present context, the SCAD estimator is given by,
\begin{eqnarray}
\hat{\beta}_{SCAD} = 
\left\{ \begin{array}{ll}
\operatorname{sgn}(\omega)(\vert \omega \vert - \lambda)_{+} &\mbox{ when $\vert \omega \vert \leq 2\lambda$} \\
\dfrac{\{ (a-1)\omega - \operatorname{sgn}(\omega)a\lambda \} }{a-2} &\mbox{ when $2\lambda < \vert \omega \vert \leq a\lambda$}\\
\omega & \mbox{ when $\vert \omega \vert >  a\lambda$}
\end{array} \right.
\label{equ: scad}
\end{eqnarray}
where $a > 2$ is an additional tuning parameter. The continuous differentiable penalty function for SCAD estimator is defined by,
\begin{eqnarray}
p^{'}_{\lambda}(\beta) = \lambda\left\{ I(\beta \leq \lambda) + \dfrac{(a\lambda - \beta)_{+}}{(a-1)\lambda}I(\beta > \lambda) \right\}  \quad 
\mbox{for $a>2$ and $\beta>0$}
\end{eqnarray} 

To sum up, both estimators are members of this penalized likelihood family. Lasso has better performance when the noise to signal ratio is large, but this approach creates bias. SCAD can generate variable selection results without generating excess biases.

\paragraph{Iterated-SIS algorithm for estimation}
SIS method is proposed by \cite{fan2008sure} to select important variables in ultra high-dimensional linear models. The proposed two-stage procedure can perform better than other methods in the sense of statistical learning problems. The SIS method is based on the concept of sure screening, is defined as the correlation learning which filters out the features that have weak correlation with the response. By sure screening, all the important variables survive after variable screening with probability tending to 1. \cite{fan2009ultrahigh} improve iterated-SIS to a general pseudo-likelihood framework by allowing feature deletion in the iterative process. \cite{fan2010sure} further extend the SIS model and consider an independent learning by ranking the maximum marginal likelihood estimator or maximum marginal likelihood itself for generalized linear models. Here we combine the VAR($p$) model and SIS algorithm to find out the key elements in a big system. The basic idea of SIS is introduced in the following.

Let $\omega = (\omega_{1}, \omega_{2}, \ldots, \omega_{p})^{\top}$ be a p-vector that is obtained by component-wise regression, i.e.,
\begin{eqnarray}
\omega=X^{\top}y
\end{eqnarray}
where $y$ is $n$ vector of response and $X$ is a $n\times p$ data matrix. $\omega$ is a vector of marginal correlations of predictors with the response of predictors with the response variable, rescaled by the standard deviation of the response. 

When there are more predictors than observation, LS (least square) estimator is noisy, that's why ridge regression is considered. Let $\omega^{\lambda} =(\omega_{1}^{\lambda}, \ldots, \omega_{p}^{\lambda})^{\top}$ be a $p-$vector obtained by ridge regression, i.e.,
    \begin{eqnarray}
    \omega^{\lambda} = (X^{\top}X + \lambda I_{p}) ^{-1}  X^{\top}y
    \end{eqnarray}
where $\lambda > 0$ is a regularization parameter. Obviously, when $\lambda \rightarrow 0$, $\omega^{\lambda} \rightarrow \hat{\beta}_{LS}$ and $\lambda \rightarrow \infty$, $\lambda \omega^{\lambda} \rightarrow \omega$. The component-wise regression is a specific case of ridge regression with  $\lambda = \infty$.\\

The iterated-SIS algorithm applied for estimating the VAR($p$) model is,
 \begin{enumerate}
\item Apply SIS for initial screening, reduce the dimensionality to a relative large scale d;
\item Apply a lower dimensional model selection method (such as lLassoasso, SCAD) to the sets of variables selected by SIS;
\item Apply SIS to the variables selected in the previous step;
\item Repeat step 2 and 3 until the set of selected variables do not decrease.
\end{enumerate}

\subsection{Connectedness measure}
We construct our network using the fashionable directional connectedness measure proposed by \cite{diebold2014network}. The connectedness is measured by cross-sectional variance decomposition, where the forecast error variance of variable is decomposed into parts attributed to the various variables in the system. 

The interactions between the variables, i.e., the directional connectedness measure $\theta_{ij}(q)$ is 
given by,
\begin{eqnarray}
\label{equ: dcm}
\theta_{ij}(q) = \dfrac{\sigma_{jj}^{-1}\sum_{q=0}^{Q-1}\left( e^{\top}_{i} \hat{B}_{q} \Sigma e_{j} \right)^{2}}{\sum_{q=0}^{Q-1}\left( e^{\top}_{i} \hat{B}_{q} \Sigma \hat{B}_{q}^{\top} e_{i} \right)}
\end{eqnarray} 
where $q$ is the lag order, $e_{i}$ is an $pK^{2} \times 1$ selection vector with unity as its $i$-th element and zeros elsewhere. $\Sigma = \operatorname{E}\left( u_{t}u_{t}^{\top} \right)$, is the covariance matrix of the non-orthogonalized VAR($p$) in equation \eqref{equ: var1}, with $\sigma_{jj}$ is the corresponding $j$-th diagonal element of $\Sigma$. $\hat{B}_{l}$ are the coefficient matrices of \eqref{equ: ma11}.

With iterated-SIS algorithm to estimate the sparse VAR structure, we can acquire its moving average (MA) transformation,
\begin{eqnarray}\label{equ: ma11}
y_{t} = \sum_{i=0}^{\infty}B_{i} u_{t-i}
\end{eqnarray} 
where the coefficient matrices $B_{i}$ obey $B_i = \sum_{j=1}^{iy}B_{i-j} A_{j}$, with $B_0 = I_{K}$ and $A_{j}=0$ for $j>p$. $A_{j}, j=1,2,\ldots, p$ is the coefficient matrices of VAR($p$) model.

To measure the persistent effect of a shock on the behavior of a series, we aim to acquire the population connectedness table  \ref{tab: conn11}, according to \cite{diebold2014network}. 

\begin{table}[h]
\scriptsize
  \centering
    \begin{tabular}{c|ccccccc}
   \hline\hline
 & $x_{1}$  & $x_{2}$  & \ldots  & $x_{n}$  & From others\\
 \hline
$x_{1}$  & $\theta_{11}(q)$ &  $\theta_{12}(q)$  & \ldots  &  $\theta_{1n}(q)$  & $\sum_{j=1}^{n}\theta_{1j}(q),  j\neq1$ \\
$x_{2}$  & $\theta_{21}(q)$ &  $\theta_{22}(q)$  & \ldots  &  $\theta_{2n}(q)$  & $\sum_{j=1}^{n}\theta_{2j}(q), j\neq2$ \\
\vdots  & \vdots  &\vdots  &   & \vdots & \vdots \\
$x_{n}$  & $\theta_{n1}(q)$ &  $\theta_{n2}(q)$  & \ldots  &  $\theta_{nn}(q)$  & $\sum_{j=1}^{n}\theta_{nj}(q), j\neq n$ \\
 \hline
To others & $\sum_{i=1}^{n}\theta_{i1}(q), i\neq1$  &$\sum_{i=1}^{n}\theta_{i2}(q), i\neq2$  & \ldots  & $\sum_{i=1}^{n}\theta_{in}(q), i\neq n$  & $\dfrac{1}{n}\sum_{i=1,j=1}^{n}\theta_{ij}(q), i\neq j$ \\
 \hline \hline
\end{tabular}
\caption{Connectedness table of interest.}\label{tab: conn11}
\end{table}

\normalsize
The rightmost column gives the "from" effect of total connectedness, and the bottom row gives the "to" effect. In particular, the directional connectedness "from" and "to" associated with the forecast error variation $\theta_{ij}$ for specific power contract when the arising shocks transmit from one asset to the others. These two connectedness estimators can be obtained by adding up the row or column elements,  the pairwise directional connectedness from $j$ to $i$ is given by,
\begin{eqnarray}
C_{i\leftarrow j} = \theta_{ij}(q)
\end{eqnarray}
The total directional connectedness "from" $C_{i\leftarrow \cdot}$ (others to $i$), "to" $C_{\cdot \leftarrow j}$ ($j$ to others) and the corresponding net connectedness are defined as 
\begin{eqnarray}
C_{i\leftarrow \bullet} &=& \sum_{j=1} ^{n}\theta_{ij}(q), i\neq j \nonumber\\
C_{\bullet \leftarrow j} &=& \sum_{i=1} ^{n}\theta_{ij}(q), i\neq j \nonumber\\
C_{i} &=& C_{to} - C_{from} = C_{\bullet \leftarrow i} - C_{i\leftarrow \bullet} 
\end{eqnarray}
The "to''  connectedness measures its total contribution to the forecast error variance of other components in the system, and therefore captures the systemic risk of the component.
\section{Data and Model Selection}\label{sec: empiricalres}
\subsection{Data sets}
As introduced in Section \ref{subsec: market}, EEX offers continuous trading data of Phelix Futures. The available load profiles are base, peak and off-peak. The available products with different maturities have five kinds: Day/Weekend Futures, Week Futures, Month Futures, Quarter Futures and Year Futures. Nevertheless the products of Day/Weekend Futures and Week Futures only have the off-peak load data, for all other contracts base and peak only. Here we recall the underlying of the Phelix Futures data, the Phelix Base contract is average price of the hours 1 to 24 for electricity traded on spot market, while the Phelix Peak is the average price of the hours 9 to 20 for electricity traded on spot market. Therefore we involve the products of spot prices as well. The contracts of spot prices are diversified in Hours from 00-01h up to 23-24h, and in Blocks of Base Monthly, off-peak 01-08, off-peak 21-24, Peak Monthly. The dataset we constructed is provided by Bloomberg, we have 90 kinds of contracts in total. The time span is from 30.09.2010 to 31.07.2015. All the contracts are listed on Table \ref{tab: contracts} with detailed information in Table \ref{tab: coninfo}. 

\begin{figure}[h]
\begin{subfigure}{.5\textwidth}
  \centering
  \includegraphics[width=1.0\linewidth]{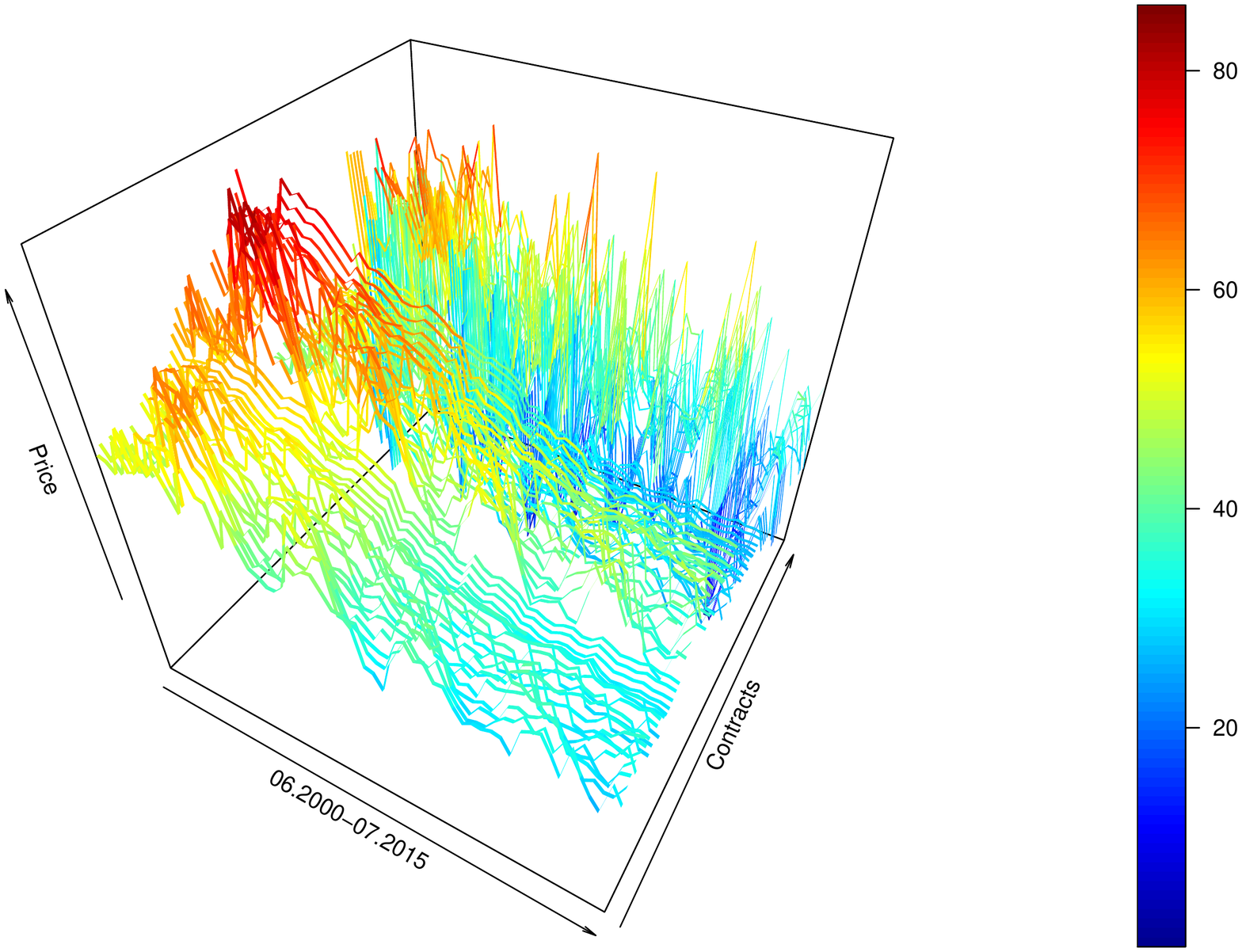}
  \caption{Ribbon plot of prices over 90 contracts}
  \label{fig: 3dplot}
\end{subfigure}%
\begin{subfigure}{.5\textwidth}
  \centering
  \includegraphics[width=.9\linewidth]{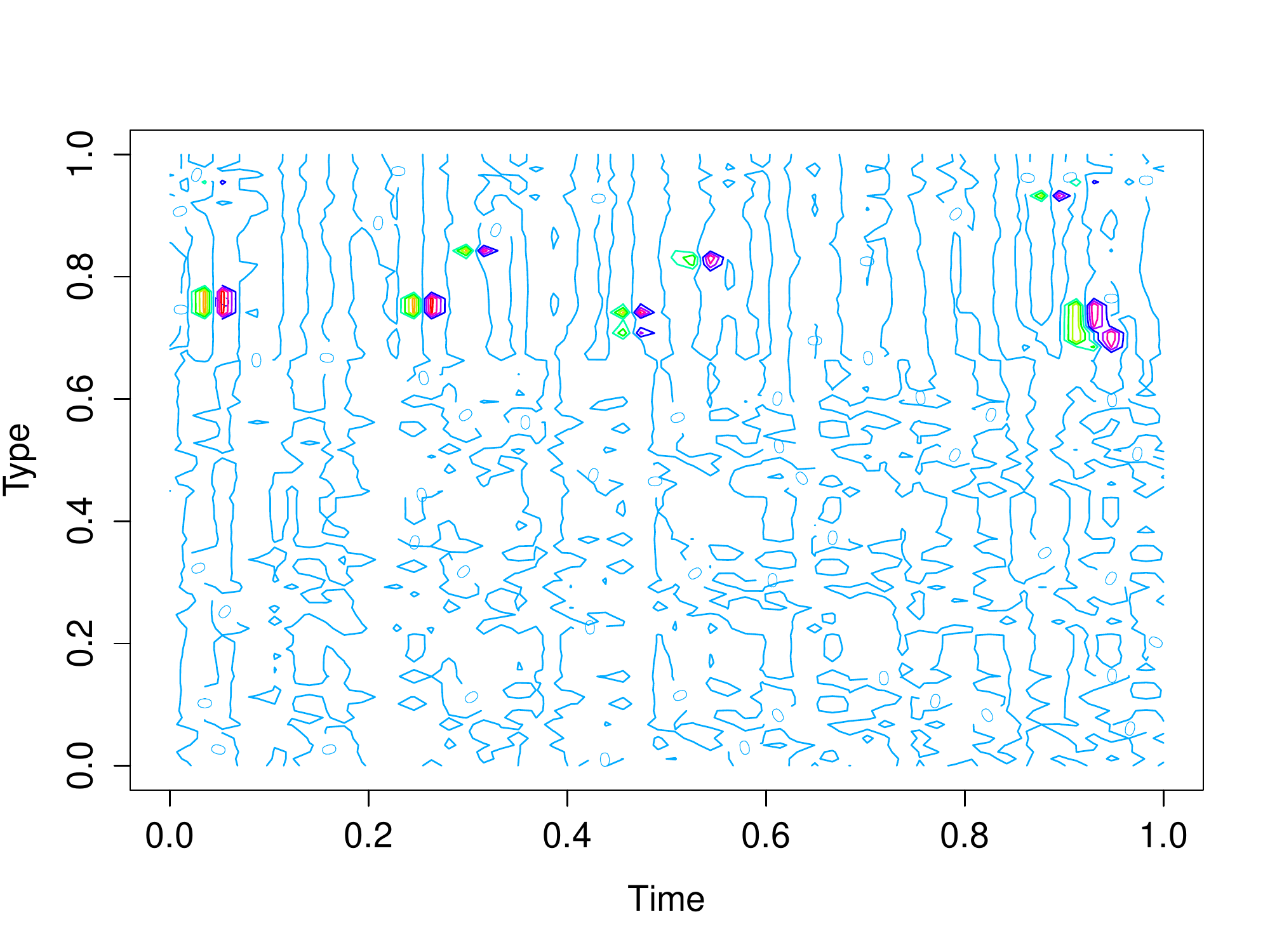}
  \caption{Contour plot of log return}
  \label{fig: contour}
\end{subfigure}
\caption{Overview of dataset}
\end{figure}

\normalsize
To remove the redundant variable, we apply screening technique to select variables using the Phelix Futures consisting of different contracts and over different maturities. To implement the VAR model, first order difference of the data in Figure \ref{fig: 3dplot} is required to transform non-stationary data to stationary time series. The contour plot of the constructed dataset is depicted in Figure \ref{fig: contour}.

\begin{landscape}
\begin{figure}[h]
\centering
\includegraphics[scale=0.55]{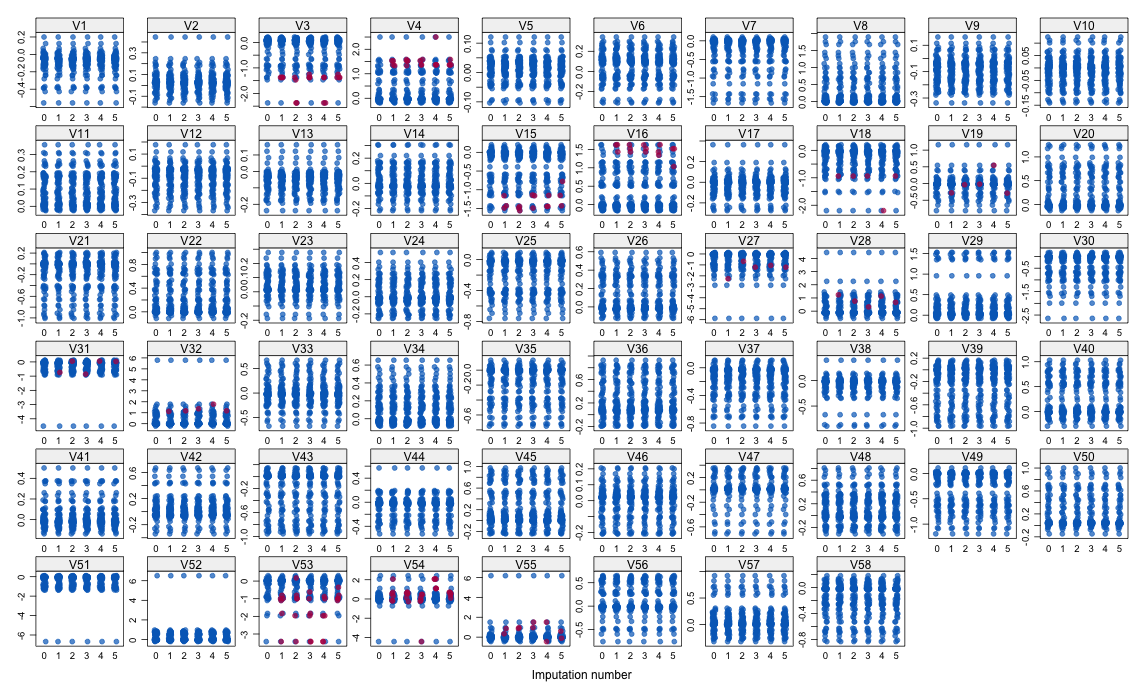}
\caption{Pattern of imputed data. \label{fig: imputdata}}
\end{figure}
\end{landscape}

In the market of Phelix Futures, final settlement at negative price is also possible. There are some missing values after transforming the original data to stationary time series by first order difference. To deal with the missing data, some quick fixes such as mean-substitution may be fine in some cases. While such simple approaches usually introduce bias into the data, for instance, applying mean substitution leaves the mean unchanged (which is desirable) but decreases variance, which may be undesirable. In our paper, we impute missing values with plausible values drawn from a distribution using an approach  proposed by \cite{van2000multivariate}. The patterns of missing data for the original dataset and imputation dataset are compared in the Figure \ref{fig: imputdata}. The distributions of the variables are shown as individual points, the imputed data for each imputed dataset is shown in magenta while the density of the observed data is shown in blue. The distributions are expected to be similar based on the assumption. We can observe that the shape of the magenta points (imputed) matches the shape of the blue ones (observed). The matching shape tells us that the imputed values are indeed plausible values. 

\subsection{Model Selection}
The purpose of this section is to compare the performance of the regularization approaches and to select the best model used to construct connectedness measure. The estimation steps are as follows,
   \begin{enumerate}
   \item Given the lag order $p$, $p$ is a constant.
   \item Recall iterated-SIS algorithm in \ref{subsec: reis}, estimate VAR($p$) using either iterated-SIS-Lasso or iterated-SIS-SCAD.
   \item Select the best model with model selection criterion given by,
   \begin{eqnarray}
   IC(p) = \log\vert \hat{H}(p) \vert + \varphi(K, p)c_{T}
   \label{equ: lagic}
   \end{eqnarray}
where $\varphi(K, p)$ is a penalty function. $c_{T}$ is a sequence indexed by the sample size $T$. The residual covariance matrix $\hat{H}(p)$ without a degrees of freedom correction is defined as, 
\begin{eqnarray}
   \hat{H}(p) = \frac{1}{T}\sum^{T}_{t=1}u_{t}^{\top}u_{t}
\end{eqnarray}
Rewrite equation \eqref{equ: lagic} with different penalty functions, the three most common information criteria are the Akaike (AIC), Schwarz-Bayesian (BIC) and Hannan-Quinn (HQ),
  \begin{eqnarray}
  AIC &=& \log\vert \hat{H}(p) \vert + \frac{2}{T} pK^{2} \\
  HQ &=& \log\vert \hat{H}(p) \vert + \frac{2\log \log T}{T} pK^{2}\\
  BIC &=& \log\vert \hat{H}(p) \vert + \frac{\log T}{T} pK^{2}
  \end{eqnarray}   
   \item The selected model will be used to construct directional connectedness $\theta_{ij}(q)$ defined in \eqref{equ: dcm}.
   \end{enumerate}

The comparison of three information criteria is reported in Table \ref{tab: modelic}. We observe that the VAR($2$) estimated by iterated-SIS-Lasso performs best, with smallest IC values given by: AIC 4.5006, HQ 4.6426 and BIC 5.6076.  The model with smaller IC values is more likely to be the true model.

\begin{table}[h]
\begin{center}
\begin{tabular}{r|rrr}
\hline\hline
Model & AIC & HQ & BIC \\
\hline
iterated-SIS-Lasso, $p=1$ & 4.5686  & 4.7249 & 5.7864 \\
iterated-SIS-Lasso, $p=2$ & 4.5006 & 4.6426 & 5.6076 \\
iterated-SIS-Lasso, $p=3$ & 6.9854 & 7.2315 & 10.0345 \\
\hline
iterated-SIS-SCAD, $p=1$ & 4.5714 & 4.7277 & 5.7892 \\
iterated-SIS-SCAD, $p=2$ & 6.1043 & 6.1043 & 9.5782 \\
iterated-SIS-SCAD, $p=3$ & 7.2559 & 7.6820 & 10.5770 \\
\hline\hline
\end{tabular}
\caption{Model selection results according to the AIC, BIC and HQ criteria.}
\label{tab: modelic}
\end{center}
\end{table}

\begin{figure}
\centering
\includegraphics[scale=0.7]{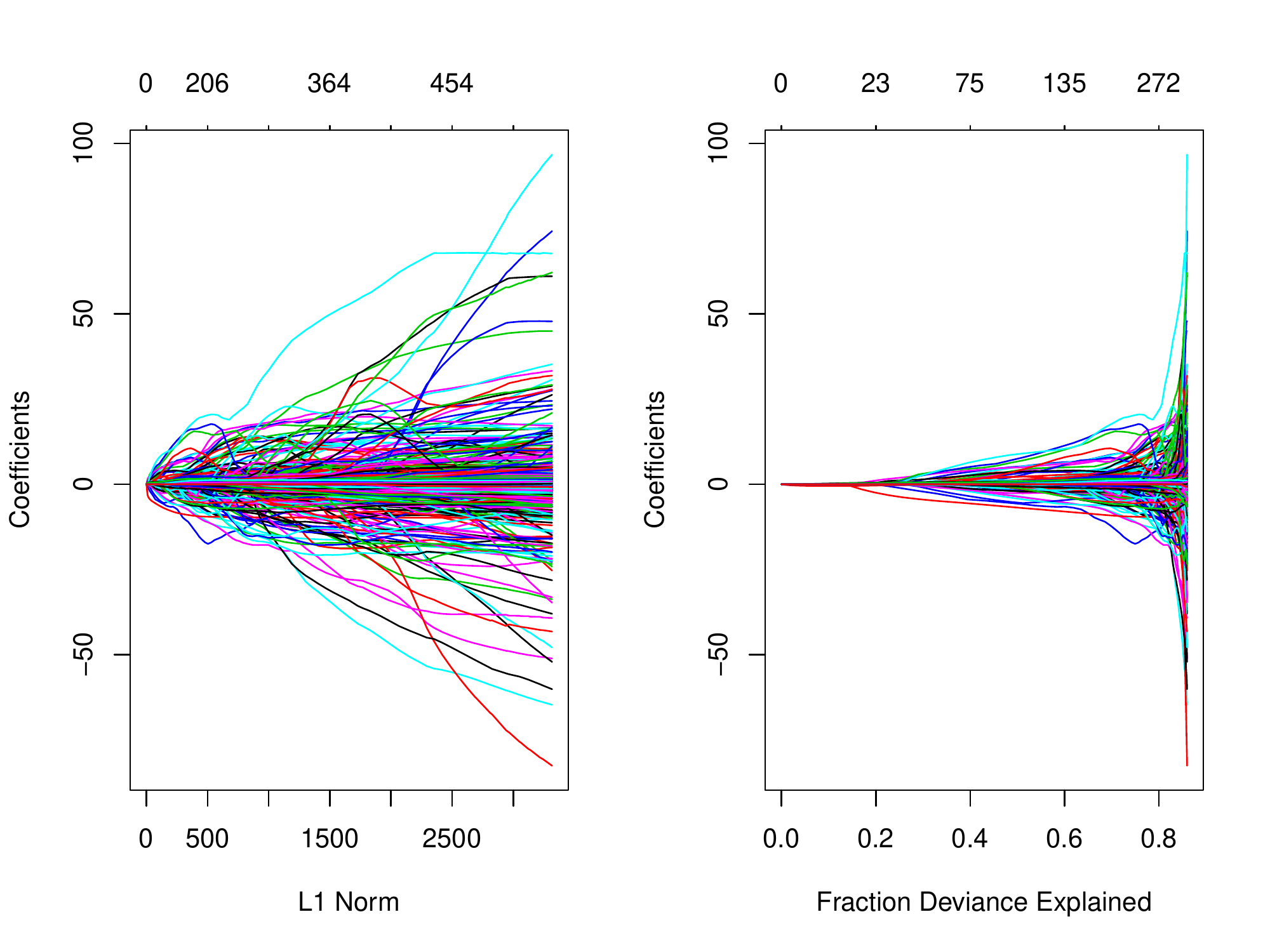}
\caption{Iterated-SIS-Lasso VAR(2) estimation results.}
 \label{fig: lassoest}
\end{figure}

We produce the coefficient paths for iterated-SIS-Lasso VAR estimation in Figure \ref{fig: lassoest}. Each curve corresponds to a variable. It shows the path of its coefficient against $\ell_{1}$-norm of the whole coefficient vector as $\lambda$ varies. 
 In addition, we partition the dataset into two samples: we select the in-sample dataset as 30.09.2010-28.11.2014, and the out-of-sample dataset used to measure model forecast performance is from 31.12.2014 to 31.07.2015. We roll each model through the out-of-sample dataset one observation at a time while each time forecasting the target variable one month ahead. By rolling window, the forecast mean squared errors (FMSE) for different models are calculated and compared in Table \ref{tab: foremse}, VAR(2) with iterated-SIS-Lasso has the smallest FMSE of 0.067, this is consistent with our previous finding. For the full sample dataset, we find that iterated-SIS-Lasso outperforms iterated-SIS-SCAD algorithm. Therefore the iterated-SIS-Lasso algorithm is selected for constructing the corresponding connectedness measure.

\begin{table}[h]
\begin{center}
\begin{tabular}{c|ccc}
\hline\hline
Lag & iterated-SIS-Lasso & iterated-SIS-SCAD \\
\hline
$p=1$ & 0.0697 & 0.0697 \\
$p=2$ & 0.0670 & 0.0701 \\
$p=3$ & 0.0923 & 0.1413 \\
\hline\hline
\end{tabular}
\caption{FMSE of out-of-sample forecasting during 31.12.2014 - 31.07.2015}
\label{tab: foremse}
\end{center}
\end{table}

\newpage
\section{Static analysis of power market connectedness}\label{sec: networkana}

\subsection{The static network across contracts}\label{subsec: static1}

\begin{figure}[h!]
\centering
\includegraphics[scale=0.55]{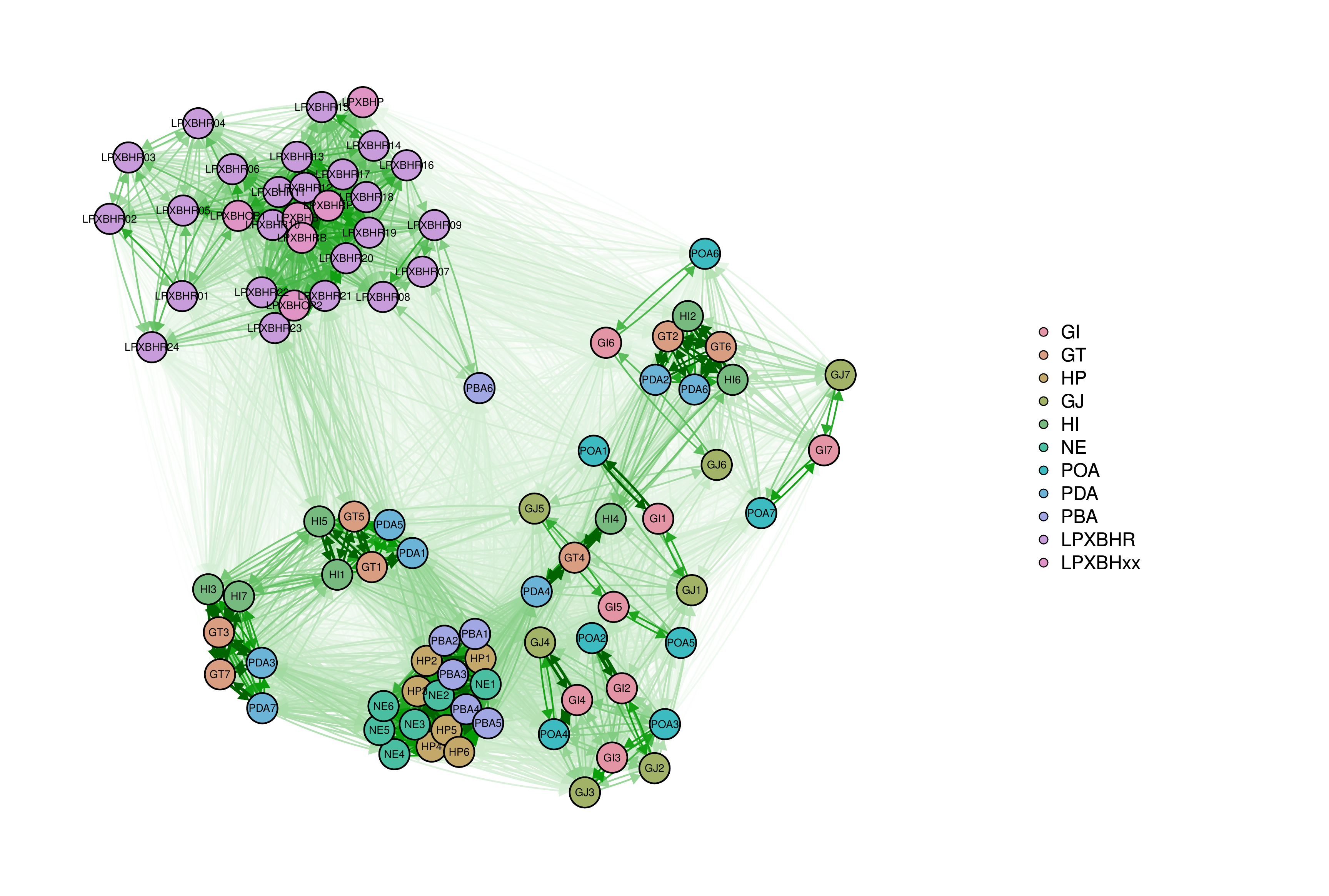}
\vspace{-1cm}
\caption{The graph for full-sample energy market network, across 11 different types with in total 90 contracts.}\label{fig: allcom}
\end{figure}
The graph of our full-sample energy market network is depicted in Figure \ref{fig: allcom}. We observe the cluster phenomena in this graph, which motivates us to study the connectedness between contracts within and across 11 different types of energy contracts. In general, the contracts that belong to the same type tend to appear inside the same cluster. We find out several pairs of strong connections between different types of contracts, 
for example, the upper-left area reveals that the LPXBHR-type and LPXBHxx-type are massively connected. In addition, a cluster consisting of HP-type (Phelix Base Year Option), NE-type (Phelix Peak Year Future) and PBA-type (Phelix Off-Peak Year Future) indicates the closer relationship among these contracts, this implies the Year derivative contracts are closer to each other while the month future and quarter future remain distinct.

\begin{table}[h]
\centering
\scriptsize
\begin{tabular}{r|rrrrrrrrrrr|r}
\hline\hline
 & GI & GT & HP & GJ & HI & NE & POA & PDA & PBA & LPXBHR & LPXBHxx & From \\ 
  \hline
GI & 1.75 & 0.46 & 0.86 & 1.46 & 0.44 & 0.71 & 1.48 & 0.48 & 0.80 & 0.15 & 0.15 & 8.74 \\ 
  GT & 0.46 & 2.33 & 0.98 & 0.58 & 2.41 & 0.93 & 0.42 & 2.03 & 0.84 & 0.30 & 0.41 & 11.70 \\ 
  HP & 0.73 & 0.84 & 5.27 & 0.63 & 0.51 & 4.78 & 0.70 & 1.21 & 4.19 & 0.06 & 0.03 & 18.95 \\ 
  GJ & 1.46 & 0.58 & 0.73 & 1.80 & 0.63 & 0.64 & 1.09 & 0.51 & 0.65 & 0.13 & 0.13 & 8.36 \\ 
  HI & 0.44 & 2.41 & 0.60 & 0.63 & 2.65 & 0.63 & 0.37 & 1.92 & 0.49 & 0.34 & 0.48 & 10.96 \\ 
  NE & 0.61 & 0.79 & 4.78 & 0.55 & 0.54 & 5.09 & 0.55 & 1.05 & 3.34 & 0.08 & 0.12 & 17.52 \\ 
  POA & 1.48 & 0.42 & 0.81 & 1.09 & 0.37 & 0.64 & 1.60 & 0.48 & 0.79 & 0.19 & 0.18 & 8.04 \\ 
  PDA & 0.48 & 2.03 & 1.42 & 0.51 & 1.92 & 1.23 & 0.48 & 1.99 & 1.27 & 0.26 & 0.32 & 11.91 \\ 
  PBA & 0.68 & 0.72 & 4.19 & 0.56 & 0.42 & 3.34 & 0.67 & 1.09 & 3.88 & 0.20 & 0.12 & 15.88 \\ 
  LPXBHR & 0.80 & 1.24 & 0.50 & 0.67 & 1.35 & 0.50 & 0.90 & 1.12 & 1.01 & 7.86 & 9.81 & \textbf{25.79} \\ 
  LPXBHxx & 0.13 & 0.35 & 0.03 & 0.11 & 0.41 & 0.12 & 0.16 & 0.28 & 0.13 & 2.70 & 3.86 & 8.28 \\ 
  \hline
  To & 9.03 & 12.18 & \textbf{20.17} & 8.59 & 11.65 & 18.62 & 8.41 & 12.16 & 17.38 & 12.28 & 15.63 & 146.12 \\ 
 Net & 0.29 &  0.48  & 1.22 &0.23 & 0.70 &1.10 & 0.37 & 0.25&  1.50 & -13.50 &7.35 \\

\hline\hline
\end{tabular}
\caption{The connectedness table for the aggregated network, including 11 types of contracts.}
\label{tab: fulltab}
\end{table}

\begin{figure}[ht]
\centering
\includegraphics[scale=0.7]{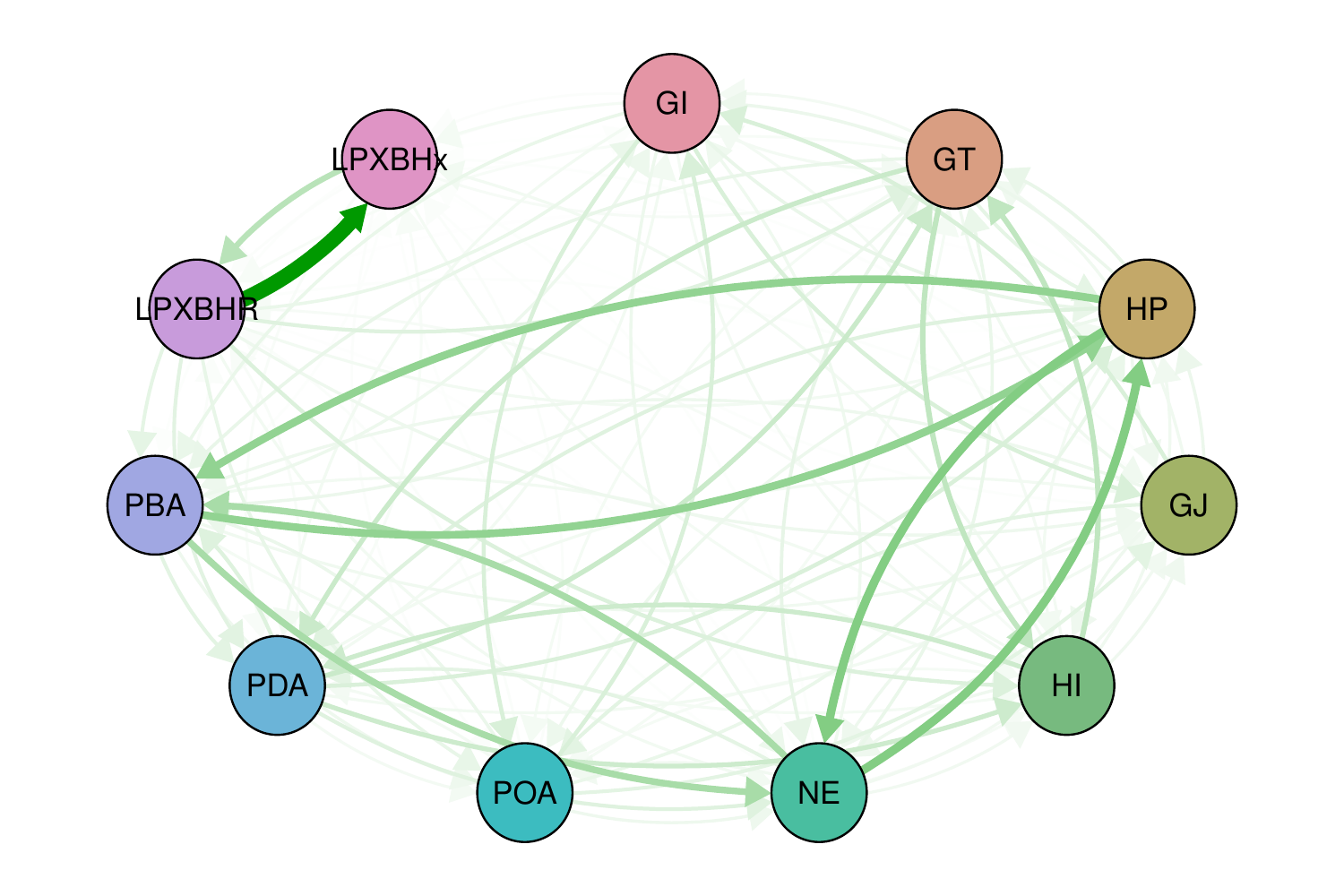}
\caption{The graph for network across 11 different contract types.}\label{fig: parcom}
\end{figure}

Given the cluster phenomena observed in the full sample connectedness graph, we aggregate the pairwise connectedness of the contracts that belong to the same type. The aggregated network is then reported in Table \ref{tab: fulltab}, with in total 11 types of contracts (more details can be found in Table \ref{tab: coninfo}). The off-diagonal elements represent the cross contractType connectedness, while within contractType connectedness are on the diagonal.

Recall that for each component in the system, the ``to'' connectedness measures its total contribution to the forecast error variance of other components in the system, making possible to capture the systemic risk of the component. In this table, it is obvious that the HP-type, NE-type and PBA-type are main contributors to systemic-wide risk. HP-type contracts have stronger links both from and to the other contract types. This is potentially interesting because, although HP-type, NE-type are important for the whole market as shown in Figure \ref{fig: parcom}, their net connectedness are negligible, with 4.52\% and 4.08\% of the total market power contracts. The reason is that these two types of contracts (HP, NE) are mutually closely interconnected, they also have ``from'' impacts to the system and thus offsetting their risk contributions. Moreover, the strongest pairwise connectedness is the impact of LPXBHxx-type on the LPXBHR-type, however the inverse impact is not significant. We also observe very strong net impacts from LPXBHxx-type on the LPXBHR-type contracts.

\subsection{Determining important market component}\label{subsec: static2}

\begin{figure}[h!]
\centering
\includegraphics[scale=0.72]{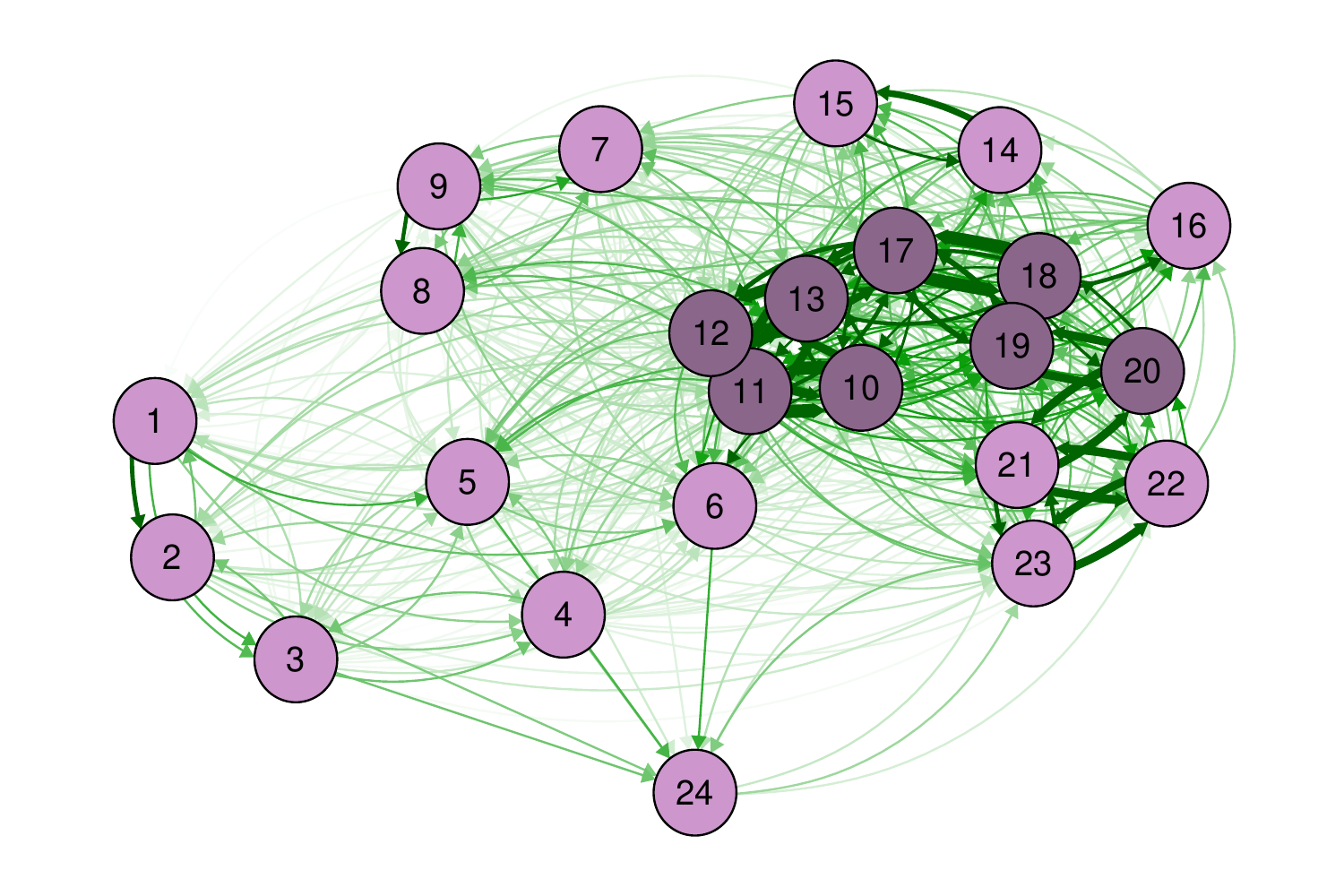}
\caption{The connectedness network graph based on different trading hours for LPXBHR-type spot contracts. The number in each node corresponds to the trading hours ranging from 00-01 to 23-24h, e.g. 12 refers to the LPXBHR12 contract with trading hours 11-12h. The LPXBHR-type  power contracts that bidding from 09-13h and 16-20h are represented with darker purple color because of their higher ``to''-connectedness.}
\label{fig: lpx24net}
\end{figure}

In terms of magnitude for different contract types reported in Table \ref{tab: fulltab}, the ``net'' directional connectedness is distributed rather tightly, in total 77.21\% of LPXBHR-type and LPXBHxx-type. We first investigate pairwise connectedness across 24 LPXBHR-type contracts (full connectedness table is available in Table \ref{tab: lptab1}).

\begin{table}[h]
\centering
\scriptsize
\begin{tabular}{c|rrr}
  \hline\hline
Power Contract & ``from''-connectedness & ``to''-connectedness & ``net''-connectedness\\ 
  \hline
LPXBHR11.Index & 12.48 & \textbf{11.08} & -1.40  \\ 
  LPXBHR10.Index & 12.27 & \textbf{10.83} & -1.45 \\ 
  LPXBHR12.Index & 11.83 & \textbf{10.55} & -1.27 \\ 
  LPXBHR13.Index & 11.55 & \textbf{10.28} & -1.27 \\ 
  \hline
  LPXBHR17.Index & 11.46 & \textbf{10.05} & -1.41 \\ 
  LPXBHR19.Index & 11.27 & \textbf{10.05} & -1.23 \\ 
  LPXBHR18.Index & 11.06 & \textbf{9.79} & -1.27 \\ 
  LPXBHR20.Index & 10.61 & \textbf{9.73} & -0.88 \\ 
  \hline
  LPXBHR21.Index & 9.51 & 9.31 & -0.19 \\ 
  LPXBHR22.Index & 8.77 & 8.51 & -0.26 \\ 
  LPXBHR23.Index & 7.80 & 8.48 & 0.68 \\ 
  LPXBHR14.Index & 7.48 & 8.16 & 0.68 \\ 
  LPXBHR16.Index & 7.02 & 7.16 & 0.14 \\ 
  LPXBHR15.Index & 6.51 & 7.15 & 0.64 \\ 
  LPXBHR01.Index & 5.99 & 6.99 & 0.99 \\ 
  LPXBHR09.Index & 5.64 & 6.54 & 0.90 \\ 
  LPXBHR06.Index & 5.60 & 6.42 & 0.83 \\ 
  LPXBHR05.Index & 5.35 & 6.30 & 0.94 \\ 
  LPXBHR04.Index & 5.28 & 6.19 & 0.91 \\ 
  LPXBHR08.Index & 5.13 & 5.65 & 0.52 \\ 
  LPXBHR07.Index & 5.04 & 4.98 & -0.06 \\ 
  LPXBHR02.Index & 4.66 & 4.87 & 0.21 \\ 
  LPXBHR03.Index & 4.36 & 4.85 & 0.49 \\ 
  LPXBHR24.Index & 2.07 & 4.84 & 2.77 \\ 
\hline\hline
\end{tabular}
\caption{Summary of ``from'', ``to'' and ``net'' connectedness for all LPXBHR-type contracts. The contracts are ranked by their system risk measure ``to''- connectedness.}
\label{tab: lptab2}
\end{table}

Figure \ref{fig: lpx24net} visualizes the connectedness network according to different trading hours, the numbered nodes correspond to the trading hours from 00-01 to 23-24h. Some blocks of high connectedness are successfully detected and represented by darker purple color, i.e. the peak spot contracts with trading hours ranging from 09-13h and 16-20h. We summarize the ``from'', ``to'' and ``net'' effects for the 24 LPXBHR-type contracts in descending order of importance, and rank the contacts by their system risk in Table \ref{tab: lptab2}. Our finding clearly shows that, the impacts from day-ahead spot power contracts that bidding between 09:00 and 13:00 have highest ``to''-connectedness values, and therefore strongest impacts on the other day-ahead spot contracts. The peak spot contracts (trading hours 09-13h, 16-20h) have very large negative ``net''-connectedness, revealing that they are main risk source in determining the electricity price.

The pairwise directional impacts between LPXBHR-type and LPXBHxx-type are plotted in Figure \ref{fig: 2lpnet}, the colors of the nodes are the same as Figure \ref{fig: allcom} (the connectedness table of the impacts from the six LPXBHxx-type contracts to the 24 LPXBHR-type spot contracts is available in the appendix, see Table \ref{tab: to24}). We find a cluster of LPXBHR10, LPXBHR11, LPXBHR12, LPXBHR13, LPXBHRP, LPXBHRB and LPXBHB. The LPXBHB (Base hours 00:00 - 14:00) contract has significant impacts on the spot contracts from hours 09 to 13, while the impacts from the LPXBHP (Peak Hours 08:00 - 20:00) contract  is negligible. In addition, both LPXBHRB (Baseload) and LPXBHRP (Peakload) contracts exhibit strong interconnectedness with the spot contracts from hours 09 - 13, however only the LPXBHRP (Peakload) contract affects the spot prices trading from hours 16 - 18. 
Moreover, the graph exhibits strong mutual links between some of the spot contracts, for example, the pairs of LPXBHR10 and  LPXBHR11, LPXBHR11 and LPXBHR12, LPXBHR12 and LPXBHR13, LPXBHR17 and LPXBHR18 and among others. 

\begin{figure}[h!]
\hspace*{-0.9cm}
\centering
\includegraphics[scale=0.8]{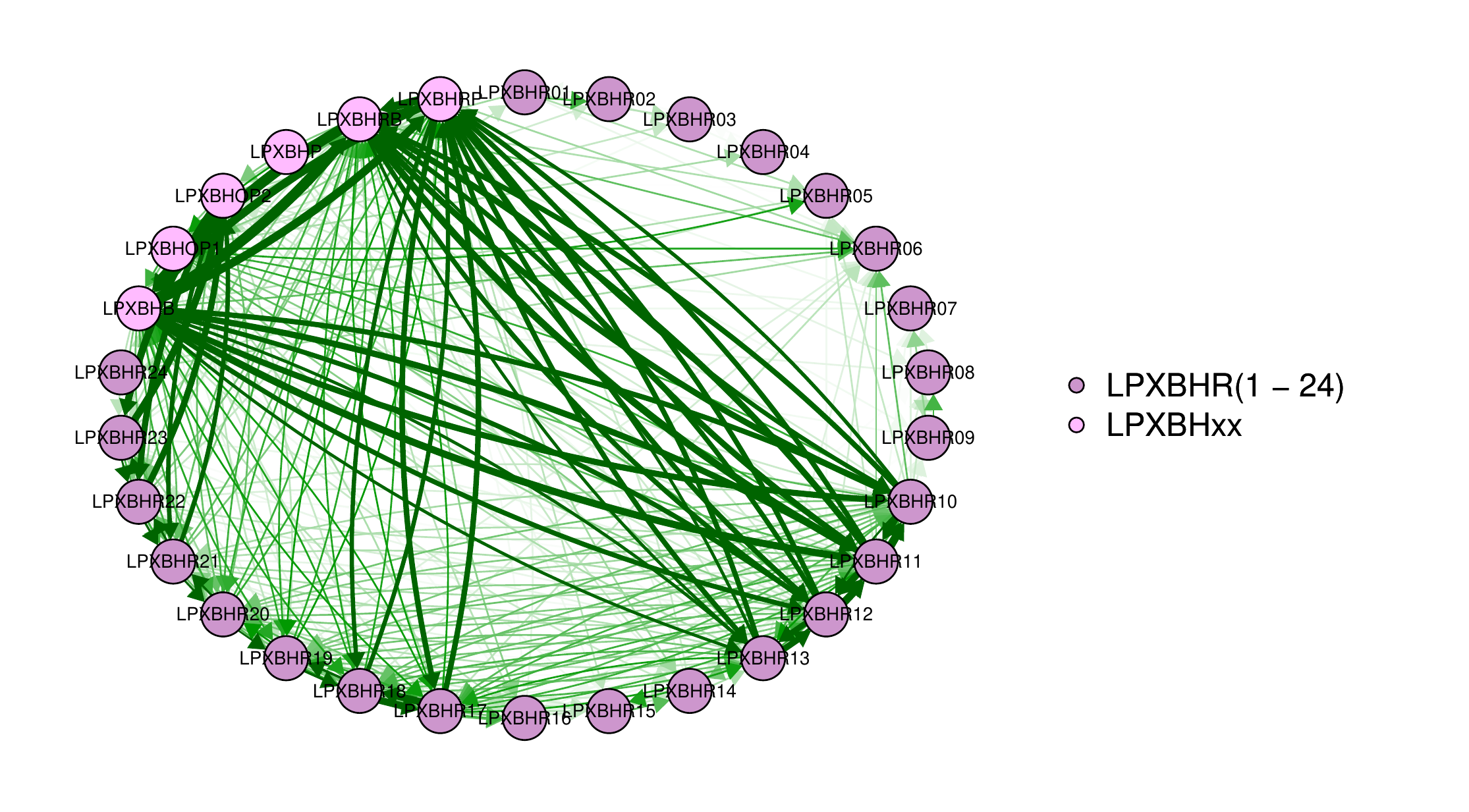}
\caption{The network graph for LPXBHR-type and LPXBHxx-type power contracts.}
\label{fig: 2lpnet}
\end{figure}

The network graph in Figure \ref{fig: hpcon} illustrates the pairwise directional connectedness between HP-type and NE-type contracts. Compared with Figure \ref{fig: allcom}, we can see that, the links between these two types are very strong. NE-type contracts are the Year Futures with maturities up to six years, the underlying of these contracts is the average price of hours 9 to 20 for electricity traded on the spot market. HP-type contracts are the European style options on the Phelix Base Future provided by EEX, the underlying of Phelix Base is the average price of the hours 1 to 24 for electricity traded on the spot market. It is calculated for all calendar days of the year as the simple average of the Auction prices for the hours 1 to 24 in the market area Germany /  Austria. This figure shows the similar connectedness pattern between HP1 and NE1 contracts.

\begin{figure}[h!]
\hspace*{0.7cm}
\centering
\includegraphics[scale=0.7]{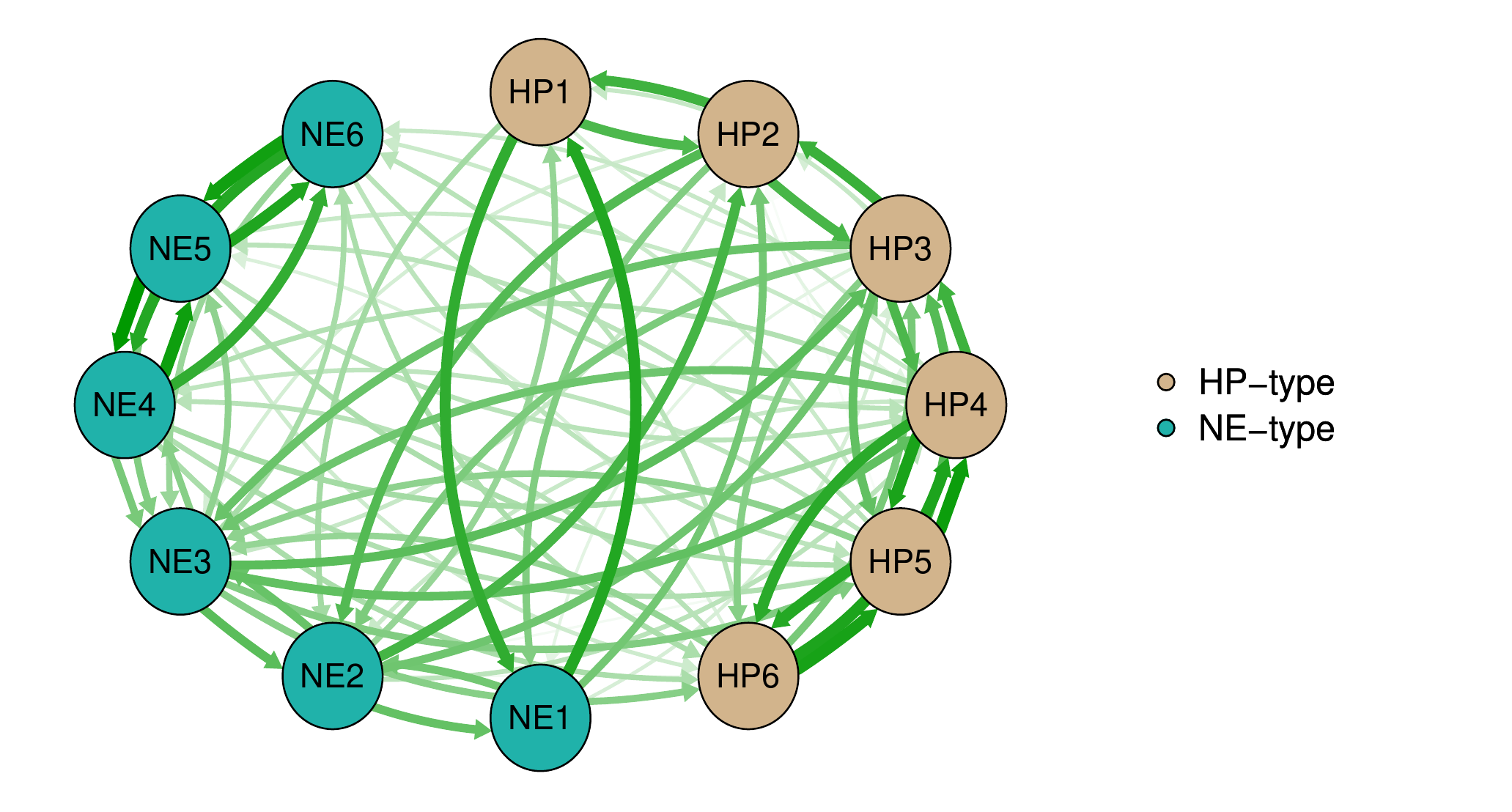}
\caption{The network graph for HP and NE-type power contracts.}
\label{fig: hpcon}
\end{figure}

\section{Dynamic analysis of power market connectedness}\label{sec: dynamicnet}
We now study the dynamic network using rolling estimation. The number of observations used in the rolling sample to compute prediction is 36 or correspondingly three years, and we examine dynamic evolution of the network for the following one year. In each window, we repeat model selection and conduct iterated-SIS algorithm to obtain the sparse estimates.

\subsection{System-wide connectedness over time}

\begin{figure}[h]
\hspace*{0.7cm}
\centering
\includegraphics[scale=0.7]{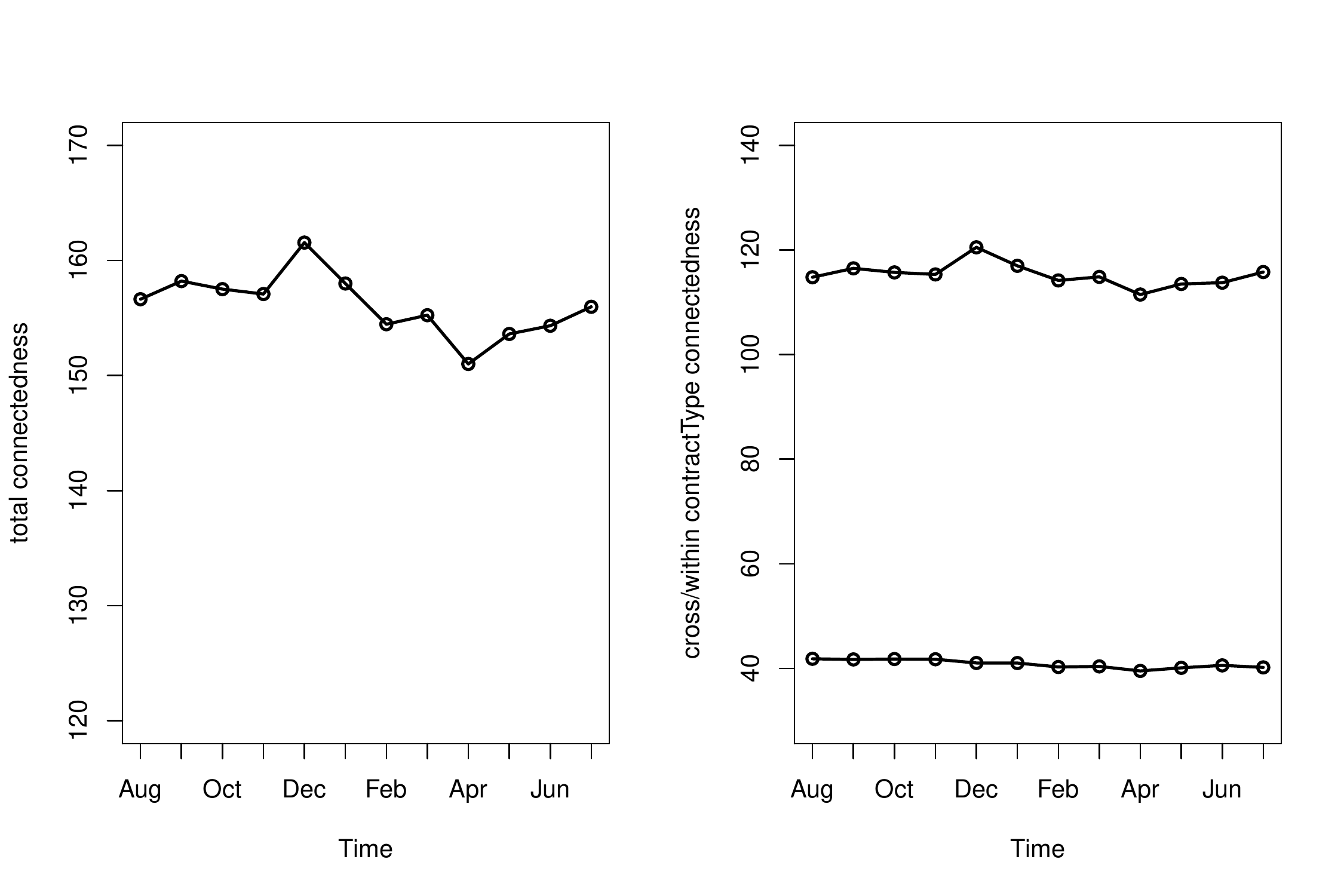}
\caption{The time-varying network for the system-wide connectedness from August 2014 to July 2015. The left panel is the time-varying full sample connectedness, it can be decomposed into two parts:  the cross contractType connectedness (upper curve in the right panel) and the within contractType connectedness(lower curve in the right panel).}
\label{fig: sysconn}
\end{figure}

We first calculate full sample system-wide connectedness for each window by summing up the total directional connectedness whether ``from'' or ``to''. In general, the full sample system-wide connectedness reflects the overall uncertainty arisen in the system. The dynamic evolution path is plotted in the left panel of Figure \ref{fig: sysconn}, with the peak of the system-wide connectedness is December. In most regions of Germany, the coldest days of the year are from around mid December to late January, which results in more work for the heating systems. In particular, the amount of electricity used for the decorations increase dramatically over the holiday season in December. After December, the value of the system-wide connectedness decreased to the lowest value of April, and it then increased gradually with increasing temperature until September. 

To measure the system-wide interaction, we further decompose the full sample system-wide connectedness into two parts: cross contractType connectedness and within contractType connectedness as shown in the right panel of Figure \ref{fig: sysconn}. The cross contractType connectedness sums up the directional connectedness between the contracts coming from different contract types, while within contractType connectedness is the sum of directional connectedness of all the contracts within the same contract type. The higher values of cross contractType connectedness (upper curve) indicate that it is the main driving force for system-wide connectedness. However the within-contractType connectedness becomes less important, its values remain around 40 for the whole period. 

We further conduct a robustness risk by computing full sample system-wide connectedness $C_{h,t}$ at different forecast horizons $h_{1}=8, h_{2}=9$. Then the dynamic networks of the resulting time-varying $C_{h_{1},t}, C_{h_{2},t}$ are compared with the above full sample system-wide connectedness $C_{h_{0},t}$, $h_{0}=10$ . To achieve this, we apply a Welch two sample t-test on $H_{0}: C_{h_{0},t} = C_{h_{1},t}$ against $H_{a}: C_{h_{0},t} \neq C_{h_{1},t}$. The p-value of the Welsh t-test is $0.8994$. We therefore cannot reject the null hypothesis, indicating that  there is not enough evidence of a difference between the (true) averages of the two groups at the usual significance level. Moreover, the p-value of the Welch two sample t-test on $H_{0}: C_{h_{0},t} = C_{h_{2},t}$ against $H_{a}: C_{h_{0},t} \neq C_{h_{2},t}$ is $0.9898$. Both large p-values suggest the robustness of dynamic network estimation. The time-varying evolution of full sample system-wide connectedness estimated at different forecast horizons can be found in Figure \ref{fig: rob1}.

\subsection{System-wide cross-contractType connectedness}

\begin{table}[ht]
\scriptsize
\centering
\begin{tabular}{c|cc|cc|cc}
  \hline   \hline
& \multicolumn{2}{c|}{``to''- connectedness}& \multicolumn{2}{c|}{``from''- connectedness}& \multicolumn{2}{c}{``net''- connectedness}\\
 & mean &s.d. & mean &s.d. & mean &s.d. \\
  \hline
GI & 9.29 (9.02 - 9.68) & 0.37 & 9.08 (8.85 - 9.33)& 0.27 & 0.22 (0.06- 0.36) & 0.16 \\ 
  GT & 12.40 (12.09 -12.57)& 0.26 & 11.92 (11.80 - 12.05)& 0.15 & 0.49 (0.24 - 0.69) & 0.19 \\ 
  HP & 21.02 (20.44 - 21.52) & 0.46 & 19.93 (19.43 - 20.41) & 0.43 & 1.09 (1.03 - 1.15) & 0.06 \\ 
  GJ & 8.49 (8.16 - 8.90) & 0.32 & 8.31 (7.99 - 8.69) & 0.32 & 0.18 (0.12 - 0.22) & 0.06 \\ 
  HI & 11.92 (11.53 - 12.20)& 0.34 & 11.16 (10.98 - 11.27) & 0.18 & 0.75(0.52 - 0.92) & 0.18 \\ 
  NE & 19.59 (19.24 - 19.92) & 0.34 & 18.49 (18.07 - 18.89) & 0.41 & 1.11 (1.02 - 1.20) & 0.09 \\ 
  POA & 8.73 (8.28 - 9.29) & 0.49 & 8.37 (8.15 - 8.70) & 0.29 & 0.36 (0.13 - 0.61) & 0.24 \\ 
  PDA & 12.35 (12.12 - 12.49) & 0.20 & 12.15 (11.95 - 12.29) & 0.19 & 0.20 (-0.03 - 0.40)& 0.20 \\ 
  PBA & 18.87 (17.61 - 20.39) & 1.35 & 17.60 (16.28 - 19.18) & 1.39 & 1.27 (1.19 - 1.33) &0.10 \\ 
  LPXBHR & 15.16 (13.62 - 16.55) & 1.46 & 29.92 (26.25 - 33.36) & 3.33 & -14.76 (-16.59 - -12.63) & 1.89 \\ 
  LPXBHxx & 18.31 (16.74 - 19.86) & 1.51 & 9.21 (8.43 - 9.85) & 0.71 & 9.10 (8.30 - 9.88) & 0.82 \\ 
   \hline \hline
\end{tabular}
\caption{The summary statistics of the connectedness measures for the aggregated network over one year, including 11 types of contracts. For each connectedness measure, we report the corresponding mean(15\% quantile - 85\% quantile) and the standard deviation.}
\label{tab: dynamic1}
\end{table}

\begin{table}[h!]
\scriptsize
\centering
\begin{tabular}{r|ccccccc}
  \hline  \hline
 & LPXBHB & LPXBHOP1 & LPXBHOP2 & LPXBHP & LPXBHRB & LPXBHRP \\ 
  \hline
  LPXBHR01.Index & 0.28 (0.01) & 0.45 (0.01) & 0.11 (0.01) & 0.15 (0.01)  & 0.28 (0.01)  & 0.19 (0.01)  \\ 
  LPXBHR02.Index & 0.30 (0.01) & 0.42 (0.02) & 0.16 (0.01) & 0.17 (0.01)  & 0.30 (0.01)  & 0.22 (0.01)  \\ 
  LPXBHR03.Index & 0.34 (0.02) & 0.54 (0.02) & 0.14 (0.01) & 0.16 (0.01)  & 0.34 (0.02)  & 0.25 (0.02)  \\ 
  LPXBHR04.Index & 0.40 (0.02) & 0.56 (0.02) & 0.17 (0.01) & 0.21 (0.01)  & 0.40 (0.02)  & 0.32 (0.01)  \\ 
  LPXBHR05.Index & 0.43 (0.02) & 0.62 (0.02) & 0.24 (0.01) & 0.26 (0.02)  & 0.43 (0.02)  & 0.34 (0.01)  \\ 
  LPXBHR06.Index & 0.51 (0.02) & 0.60 (0.03) & 0.33 (0.01) & 0.21 (0.01)  & 0.51 (0.02)  & 0.43 (0.02)  \\ 
  LPXBHR07.Index & 0.42 (0.01) & 0.38 (0.01) & 0.29 (0.01) & 0.11 (0.01)  & 0.42 (0.01)  & 0.38 (0.01)  \\ 
  LPXBHR08.Index & 0.43 (0.01) & 0.37 (0.01) & 0.34 (0.01) & 0.10 (0.01)  & 0.43 (0.01)  & 0.39 (0.01)  \\ 
  LPXBHR09.Index & 0.61 (0.02) & 0.49 (0.02) & 0.34 (0.01) & 0.14 (0.01)  & 0.61 (0.02)  & 0.58 (0.02)  \\ 
  LPXBHR10.Index & \textbf{0.86} (0.02) & 0.68 (0.02) & 0.47 (0.01) & 0.38 (0.01)  & \textbf{0.86} (0.02)  & \textbf{0.82} (0.02)  \\ 
  LPXBHR11.Index & \textbf{0.85} (0.02) & 0.66 (0.02) & 0.44 (0.01) & 0.45 (0.01)  & \textbf{0.85} (0.02)  & \textbf{0.83} (0.02)  \\ 
  LPXBHR12.Index & \textbf{0.82} (0.02) & 0.60 (0.02) & 0.40 (0.01) & 0.50 (0.01)  & \textbf{0.82} (0.02)  & \textbf{0.82} (0.02)  \\ 
  LPXBHR13.Index & 0.78 (0.02) & 0.53 (0.02) & 0.34 (0.01) & 0.51 (0.01)  & 0.78 (0.02)  & \textbf{0.82} (0.02)  \\ 
  LPXBHR14.Index & 0.51 (0.01) & 0.40 (0.01) & 0.17 (0.01) & 0.37 (0.01)  & 0.51 (0.01)  & 0.56 (0.02)  \\ 
  LPXBHR15.Index & 0.46 (0.01) & 0.38 (0.01) & 0.16 (0.01) & 0.30 (0.01)  & 0.46 (0.01)  & 0.50 (0.02)  \\ 
  LPXBHR16.Index & 0.56 (0.02) & 0.34 (0.01) & 0.29 (0.01) & 0.22 (0.01)  & 0.56 (0.02)  & 0.61 (0.02)  \\ 
  LPXBHR17.Index & 0.71 (0.02) & 0.44 (0.01) & 0.28 (0.01) & 0.29 (0.01)  & 0.71 (0.02)  & \textbf{0.81} (0.02)  \\ 
  LPXBHR18.Index & 0.71 (0.02) & 0.44 (0.01) & 0.32 (0.01) & 0.28 (0.01)  & 0.71 (0.02)  & \textbf{0.80} (0.02)  \\ 
  LPXBHR19.Index & 0.70 (0.02) & 0.43 (0.01) & 0.45 (0.01) & 0.34 (0.01)  & 0.70 (0.02)  & 0.73 (0.02)  \\ 
  LPXBHR20.Index & 0.67 (0.02) & 0.40 (0.01) & 0.60 (0.02) & 0.33 (0.01)  & 0.67 (0.02)  & 0.66 (0.02)  \\ 
  LPXBHR21.Index & 0.58 (0.02) & 0.33 (0.01) & 0.78 (0.02) & 0.29 (0.01)  & 0.58 (0.02)  & 0.53 (0.01)  \\ 
  LPXBHR22.Index & 0.49 (0.01) & 0.31 (0.01) & \textbf{0.84} (0.02) & 0.18 (0.01)  & 0.49 (0.01)  & 0.39 (0.01)  \\ 
  LPXBHR23.Index & 0.43 (0.01) & 0.22 (0.01) & \textbf{0.84} (0.02) & 0.33 (0.01)  & 0.43 (0.01)  & 0.36 (0.01)  \\ 
  LPXBHR24.Index & 0.22 (0.01) & 0.12 (0.01) & 0.57 (0.02) & 0.16 (0.01)  & 0.22 (0.01)  & 0.18 (0.01) \\ 
   \hline\hline
\end{tabular}
\caption{The connectedness table reflects how the LPXBHxx-type contracts may affect spot contract according to different trading hours. In this table we report the mean of the time-varying networks, together with the s.d. inside the parentheses. Each element measures the directional pairwise connectedness from the $j$th contract of LPXBHxx-type to the $i$th LPXBHR-type, i.e. $C_{LPXBHR_{i} \leftarrow LPXBHxx_{j}}$. The numbers larger than 0.8 are marked with bold font.} \label{tab: dynamictab2}
\end{table}

To investigate the cross contractType connectedness, we aggregate the pairwise connectedness of the contracts that belong to the same type. The resulting time-varying networks across different contract types are summarized in Table \ref{tab: dynamic1}. For each contract type, we report the mean and standard deviation (s.d.), the numbers inside the parentheses represent the 15\% and 85\% quantiles respectively. In general, the results are consistent with the findings in section \ref{subsec: static1}. The HP-type, NE-type and PBA-type contracts have relatively higher values of both ``to''-connectedness and ``from''-connectedness, but their net impacts in the system are very low. Moreover, the LPXBHxx-type contracts have significant net impacts on the LPXBHR-type contracts, similar results are discussed in section \ref{subsec: static2}. 

According to different trading hours, we further discuss how LPXBHxx-type contracts may affect the LPXBHR-type contracts traded in the power derivative market. The mean and s.d. for the time-varying connectedness $C_{LPXBHR_{i} \leftarrow LPXBHxx_{j}}$ are reported in Table \ref{tab: dynamictab2}. The spot contracts based on bid hours from 09-13 are closely related to the contracts of LPXBHB, LPXBHRB and LPXBHRP. We also find that the LPXBHRP affects the spot contracts from hours 16-19h most. To control for a relatively stable Germany spot electricity price, the risk transmission channels among the contracts are not negligible. Investments in LPXBHB, LPXBHRB and LPXBHRP contracts help to limit the potential risk of loss when there are adverse movements of spot prices. This results may provide guide for policy maker, energy companies and investors.

We also compute the averaged pairwise connectedness among the spot contracts based on different trading hours, the interaction between the LPXBHR-type contracts is depicted in Figure \ref{fig: dynamic2}. We observe strong pairwise interconnection between the neighboring contracts. For example, the LPXBHR12 has stronger connection with its nearest two neighbors LPXBHR11 and LPXBHR13, but contract LPXBHR13 does not have strong linkage with LPXBHR14. The most influential contracts are identified as the contracts based on trading hours from 09-13 and 16-20. This is potentially interesting as it provides pricing signals affecting the electricity trading.

\begin{figure}[h!]
\hspace*{0.7cm}
\centering
\includegraphics[scale=0.7]{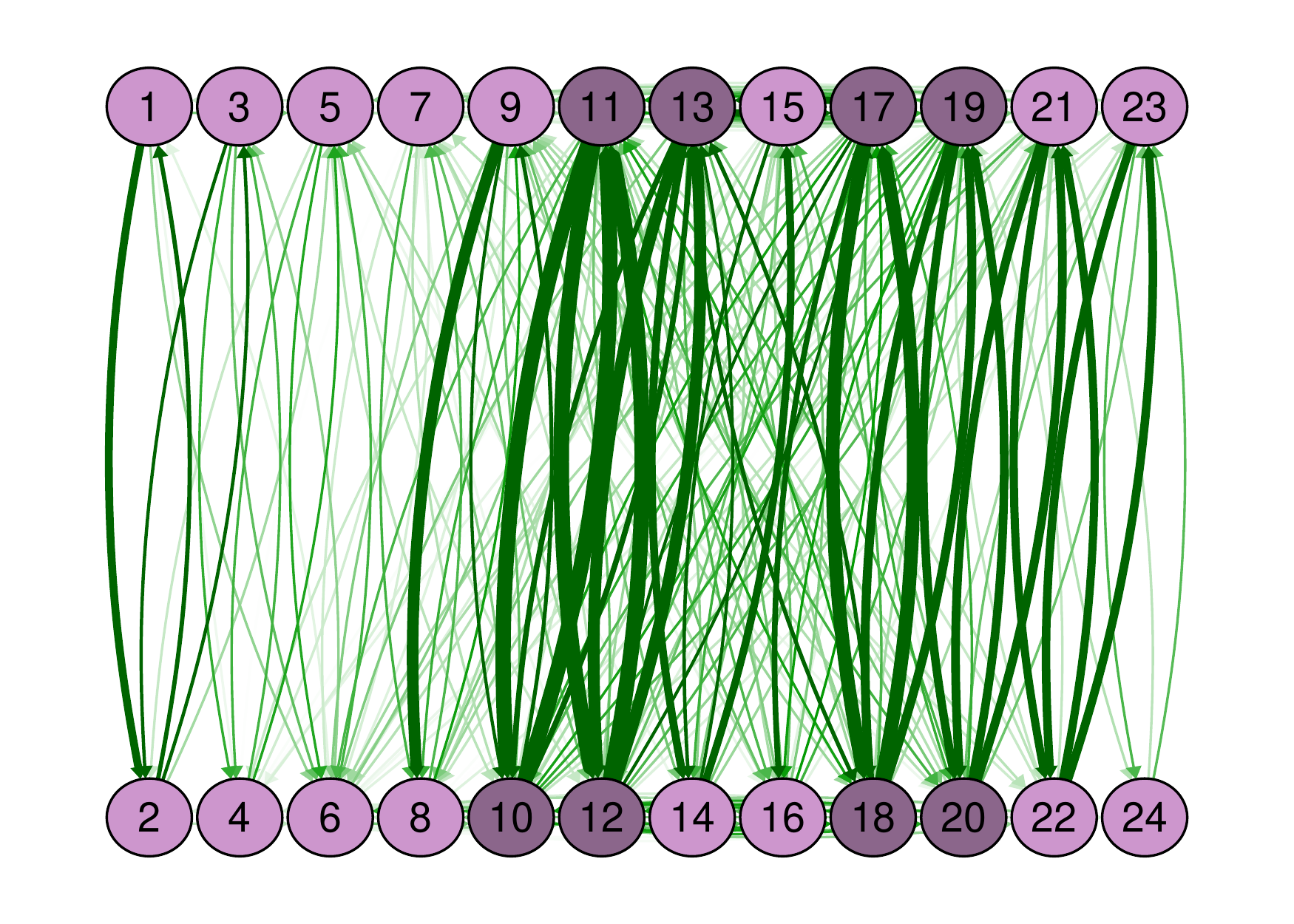}
\caption{The averaged pairwise connectedness among the LPXBHR-type contracts over the rolling period. The numbered nodes correspond to the underlying trading hours for each contract. The colors are the same as Figure \ref{fig: lpx24net}, the contracts with higher connectedness values are represented with darker purple color.}
\label{fig: dynamic2}
\end{figure}

\section{Conclusion}\label{sec: conc}

This paper combines regularization approach with dynamic network analysis in an ultra high-dimensional setting. We empirically analyze the sparse interconnectedness of the full German power derivative market, clearly identify the significant channels of relevant potential risk spillovers and thus complement the full picture of system risk. As we know, electricity is not storable and may be affected by various factors on the supply and demand side of the market, such as policy changes, weather conditions and external economic uncertainties. Nowadays Germany is transforming its power system towards renewable energy, analysis of German power derivative market thus provides useful insights for power generation companies and transmission organizations across the globe. 

Our empirical findings suggest that the Phelix base year options and peak year futures are the main contributors to the system risk. However these two types of contracts are mutually closely interconnected, they also have high ``from'' impacts received from the system and thus offsetting their risk contributions. In addition, the connectedness between different contract types are more significant compared with the connectedness among contracts within the same type. Therefore it is important for policy makers and investors to take the interdependence between different contractTypes into account. Other interesting conclusions arise from the connectedness of spot and future contracts, when we examine each contract according to different trading hours. For spot contracts, we observe strong pairwise interconnections between the neighboring contracts especially for the contracts trading in the peak hours, this provides investors pricing signals affecting the electricity trading. In addition, the monthly base future (BHB), monthly peakload(BHRP)/baseload(BHRB) are identified as main driving force for the peak-hour spot contracts. This evidence has implications for regulators to control for a relatively stable Germany spot electricity price. For energy companies and investors, it is important to diversity their existing portfolio rather than having large holdings of individual electricity contract, for example, investments in LPXBHB, LPXBHRB and LPXBHRP contracts help to limit the potential risk of loss when there are adverse movements of spot prices. Another important characteristic of electricity is seasonality, this characteristic is reflected in our dynamic network analysis. We observe the dynamic evolution of full-sample system-wide connectedness in case of weather condition and external uncertainty, for example the full-sample system-wide connectedness increased gradually with increasing temperature from April until September. In general, with the wide range of power derivative contracts trading in the German electricity market, we are able to identify, estimated the risk contribution of individual power contract, this helps us to have a better understanding of the German power market functioning and environment.


\bibliographystyle{apalike}
\bibliography{mybib}

\newpage
\appendix
\section{Appendix} \label{app}

\begin{table}[h!]
\begin{small}
\begin{center}
\scriptsize
\begin{tabular}{r|lllll}
\hline  \hline
Symbol & &Types& \\
\hline
GI & GI1.Comdty & GI2.Comdty & GI3.Comdty & GI4.Comdty \\
 & GI5.Comdty & GI6.Comdty & GI7.Comdty &&\\
\hline
GT & GT1.Comdty & GT2.Comdty & GT3.Comdty & GT4.Comdty \\
& GT5.Comdty &GT6.Comdty & GT7.Comdty && \\ 
\hline
HP & HP1.Comdty & HP2.Comdty & HP3.Comdty & HP4.Comdty\\
 & HP5.Comdty & HP6.Comdty & &  \\ 
 \hline
GJ & GJ1.Comdty & GJ2.Comdty & GJ3.Comdty & GJ4.Comdty\\
  & GJ5.Comdty & GJ6.Comdty & GJ7.Comdty &  \\ 
 \hline 
HI & HI1.Comdty & HI2.Comdty & HI3.Comdty & HI4.Comdty \\
& HI5.Comdty & HI6.Comdty & HI7.Comdty &  \\ 
  \hline
NE & NE1.Comdty & NE2.Comdty & NE3.Comdty & NE4.Comdty \\
& NE5.Comdty & NE6.Comdty &  &  \\
  \hline
POA & POA1.Comdty & POA2.Comdty & POA3.Comdty & POA4.Comdty\\
  & POA5.Comdty & POA6.Comdty & POA7.Comdty &  \\ 
  \hline
PDA & PDA1.Comdty & PDA2.Comdty & PDA3.Comdty & PDA4.Comdty \\
& PDA5.Comdty & PDA6.Comdty & PDA7.Comdty &  \\ 
  \hline
PBA & PBA1.Comdty & PBA2.Comdty & PBA3.Comdty & PBA4.Comdty \\
& PBA5.Comdty & PBA6.Comdty & & \\ 
  \hline
LPXBHR & LPXBHR01.Index & LPXBHR02.Index & LPXBHR03.Index & LPXBHR04.Index\\
 & LPXBHR05.Index & LPXBHR06.Index & LPXBHR07.Index & LPXBHR08.Index\\
 & LPXBHR09.Index & LPXBHR10.Index & LPXBHR11.Index & LPXBHR12.Index \\
& LPXBHR13.Index & LPXBHR14.Index & LPXBHR15.Index  & LPXBHR16.Index \\
&LPXBHR17.Index & LPXBHR18.Index & LPXBHR19.Index & LPXBHR20.Index\\
 & LPXBHR21.Index & LPXBHR22.Index & LPXBHR23.Index & LPXBHR24.Index \\ 
  \hline
LPXBHxx & LPXBHBMI.Index & LPXBHOP1.Index & LPXBHOP2.Index & LPXBHPMI.Index \\
& LPXBHRBS.Index & LPXBHRPK.Index &  &\\ 
   \hline\hline
\end{tabular}
\caption{Phelix Futures data traded at EEX.}
\label{tab: contracts}
\end{center}
\end{small}
\end{table}

\newpage

\scriptsize
\begin{longtable}{p{.05\linewidth}|p{.4\linewidth}p{.4\linewidth}}
\hline\hline
No. &  Symbol & Description \\
\hline
1& GI1.Comdty - GI7.Comdty & Phelix Base Month Option, and the respective next six delivery months \\
\hline
2& GT1.Comdty - GT7.Comdty & Phelix Base Quarter Option, and the respective next six delivery quarters \\
\hline
3& HP1.Comdty - HP6.Comdty & Phelix Base Year Option, and the respective next five delivery years \\
\hline
4&GJ1.Comdty - GJ7.Comdty & Phelix Peak Month Future, and the respective next six delivery months \\
\hline
5&HI1.Comdty - HI7.Comdty & Phelix Peak Quarter Future, and the respective next six delivery quarters \\
\hline
6&NE1.Comdty - NE6.Comdty & Phelix Peak Year Future, and the respective next five delivery years 
\\
\hline
7&POA1.Comdty - POA7.Comdty & Phelix Off-Peak Month Future, and the respective next six delivery months \\
\hline
8&PDA1.Comdty - PDA7.Comdty & Phelix Off-Peak Quarter Future, and the respective next six delivery quarters \\
\hline
9&PBA1.Comdty - PBA6.Comdty & Phelix Off-Peak Year Future, and the respective next five delivery years \\
\hline
10&LPXBHR01.Index - LPXBHR24.Index & EEX Day-ahead Spot Market with Bid Type from 00-01 to 23-24h, e.g. LPXBHR14.Index is EEX Day-ahead Spot price based on bid hours from 13 -14.\\
\hline
11& LPXBHRxx.Index & EEX Day-ahead Spot Market with different Bid Types: LPXBHB.Index is Base Monthly 00-14h; LPXBHOP1.Index is Off Peak1 01-08h; LPXBHOP2.Index is Off Peak2 21-24h; LPXBHP.Index is Peak Monthly 08 - 20h;
 LPXBHRB.Index is Baseload; LPXBHRP.Index is Peakload.\\
   \hline\hline
\caption{Selected contracts from the file "Products 2016" provided by \href{https://www.eex.com/en/}{European Energy Exchange EEX AG}. }
\label{tab: coninfo}
\end{longtable}

\newpage
\scriptsize
\begin{longtable}{p{.1\linewidth}|p{.08\linewidth}p{.08\linewidth}p{.08\linewidth}p{.08\linewidth}p{.08\linewidth}p{.08\linewidth}p{.08\linewidth}p{.08\linewidth}}
\hline\hline
 & LPXBHR01& LPXBHR02 & LPXBHR03 & LPXBHR04 & LPXBHR05 & LPXBHR06 & LPXBHR07 & LPXBHR08\\ 
  \hline
LPXBHR01 & 1.00 & 0.71 & 0.52 & 0.35 & 0.53 & 0.43 & 0.13 & 0.14 \\ 
LPXBHR02 & 0.42 & 0.59 & 0.47 & 0.36 & 0.31 & 0.18 & 0.06 & 0.07 \\ 
LPXBHR03 & 0.27 & 0.41 & 0.54 & 0.42 & 0.36 & 0.23 & 0.12 & 0.12 \\ 
LPXBHR04 & 0.19 & 0.32 & 0.41 & 0.53 & 0.38 & 0.25 & 0.21 & 0.15 \\ 
LPXBHR05 & 0.25 & 0.24 & 0.30 & 0.33 & 0.47 & 0.37 & 0.19 & 0.15 \\ 
LPXBHR06 & 0.18 & 0.12 & 0.18 & 0.18 & 0.34 & 0.41 & 0.22 & 0.23 \\ 
LPXBHR07 & 0.10 & 0.08 & 0.13 & 0.20 & 0.22 & 0.29 & 0.45 & 0.40 \\ 
LPXBHR08 & 0.10 & 0.08 & 0.13 & 0.14 & 0.18 & 0.30 & 0.45 & 0.54 \\ 
LPXBHR09 & 0.11 & 0.08 & 0.14 & 0.15 & 0.18 & 0.35 & 0.58 & 0.70 \\ 
LPXBHR10 & 0.25 & 0.23 & 0.23 & 0.33 & 0.49 & 0.66 & 0.46 & 0.49 \\ 
LPXBHR11 & 0.27 & 0.28 & 0.26 & 0.35 & 0.50 & 0.61 & 0.42 & 0.46 \\ 
  LPXBHR12 & 0.26 & 0.29 & 0.23 & 0.30 & 0.42 & 0.48 & 0.28 & 0.31 \\ 
  LPXBHR13 & 0.23 & 0.30 & 0.24 & 0.30 & 0.39 & 0.45 & 0.21 & 0.27 \\ 
  LPXBHR14 & 0.16 & 0.15 & 0.17 & 0.23 & 0.26 & 0.30 & 0.13 & 0.13 \\ 
  LPXBHR15 & 0.22 & 0.18 & 0.17 & 0.19 & 0.22 & 0.27 & 0.14 & 0.13 \\ 
  LPXBHR16 & 0.07 & 0.06 & 0.10 & 0.13 & 0.19 & 0.34 & 0.36 & 0.41 \\ 
  LPXBHR17 & 0.13 & 0.13 & 0.18 & 0.22 & 0.35 & 0.51 & 0.29 & 0.36 \\ 
  LPXBHR18 & 0.13 & 0.11 & 0.15 & 0.20 & 0.32 & 0.47 & 0.24 & 0.27 \\ 
  LPXBHR19 & 0.15 & 0.11 & 0.12 & 0.24 & 0.32 & 0.45 & 0.33 & 0.30 \\ 
  LPXBHR20 & 0.10 & 0.07 & 0.08 & 0.17 & 0.24 & 0.38 & 0.32 & 0.28 \\ 
  LPXBHR21 & 0.06 & 0.07 & 0.05 & 0.12 & 0.17 & 0.28 & 0.27 & 0.25 \\ 
  LPXBHR22& 0.11 & 0.13 & 0.11 & 0.12 & 0.19 & 0.31 & 0.23 & 0.26 \\ 
  LPXBHR23 & 0.10 & 0.12 & 0.06 & 0.07 & 0.11 & 0.18 & 0.12 & 0.14 \\ 
  LPXBHR24 & 0.00 & 0.00 & 0.01 & 0.01 & 0.01 & 0.00 & 0.00 & 0.00 \\ 
   \hline
    & LPXBHR09 & LPXBHR10 & LPXBHR11 & LPXBHR12 & LPXBHR13 & LPXBHR14 & LPXBHR15 & LPXBHR16\\ 
  \hline
LPXBHR01 & 0.16 & 0.25 & 0.27 & 0.26 & 0.23 & 0.13 & 0.08 & 0.03 \\ 
  LPXBHR02 & 0.07 & 0.15 & 0.19 & 0.20 & 0.21 & 0.07 & 0.04 & 0.03 \\ 
  LPXBHR03 & 0.10 & 0.13 & 0.14 & 0.13 & 0.14 & 0.07 & 0.05 & 0.05 \\ 
  LPXBHR04 & 0.12 & 0.21 & 0.23 & 0.20 & 0.20 & 0.11 & 0.10 & 0.15 \\ 
  LPXBHR05 & 0.12 & 0.24 & 0.25 & 0.22 & 0.21 & 0.11 & 0.08 & 0.11 \\ 
  LPXBHR06 & 0.20 & 0.29 & 0.27 & 0.23 & 0.22 & 0.13 & 0.12 & 0.15 \\ 
  LPXBHR07 & 0.38 & 0.27 & 0.26 & 0.20 & 0.16 & 0.09 & 0.09 & 0.16 \\ 
  LPXBHR08 & 0.53 & 0.28 & 0.27 & 0.19 & 0.17 & 0.08 & 0.08 & 0.09 \\ 
  LPXBHR09 & 0.73 & 0.33 & 0.31 & 0.22 & 0.19 & 0.10 & 0.11 & 0.09 \\ 
  LPXBHR10 & 0.45 &  \textbf{0.98} &  \textbf{0.94} &  \textbf{0.82} &  \textbf{0.70} & 0.43 & 0.37 & 0.43 \\ 
  LPXBHR11 & 0.43 &  \textbf{0.94} &  \textbf{0.98} &  \textbf{0.92} &  \textbf{0.81} & 0.44 & 0.37 & 0.43 \\ 
  LPXBHR12 & 0.31 &  \textbf{0.84} &  \textbf{0.94} &  \textbf{1.00} &  \textbf{0.90} & 0.46 & 0.36 & 0.44 \\ 
  LPXBHR13 & 0.27 &  \textbf{0.71} &  \textbf{0.83} &  \textbf{0.90} &  \textbf{1.00} & 0.62 & 0.51 & 0.43 \\ 
  LPXBHR14 & 0.13 & 0.39 & 0.41 & 0.41 & 0.54 & 0.83 & 0.77 & 0.27 \\ 
  LPXBHR15 & 0.14 & 0.31 & 0.32 & 0.30 & 0.40 & 0.67 & 0.71 & 0.27 \\ 
  LPXBHR16 & 0.41 & 0.39 & 0.38 & 0.34 & 0.32 & 0.22 & 0.26 & 0.59 \\ 
  LPXBHR17 & 0.33 &  \textbf{0.68} &  \textbf{0.72} &  \textbf{0.72} &  \textbf{0.78} & 0.50 & 0.51 & 0.65 \\ 
  LPXBHR18 & 0.24 &  \textbf{0.66} &  \textbf{0.66} &  \textbf{0.65} &  \textbf{0.70} & 0.48 & 0.50 & 0.71 \\ 
  LPXBHR19 & 0.27 &  \textbf{0.63} &  \textbf{0.63} &  \textbf{0.60} & \textbf{0.63} & 0.48 & 0.47 & 0.62 \\ 
  LPXBHR20 & 0.26 & \textbf{0.64} & \textbf{0.61} & \textbf{0.58} & \textbf{0.52} & 0.34 & 0.32 & 0.49 \\ 
  LPXBHR21 & 0.24 & 0.56 & 0.54 & 0.53 & 0.44 & 0.25 & 0.22 & 0.35 \\ 
  LPXBHR22 & 0.24 & 0.51 & 0.47 & 0.42 & 0.35 & 0.17 & 0.14 & 0.32 \\ 
  LPXBHR23 & 0.14 & 0.40 & 0.41 & 0.44 & 0.39 & 0.22 & 0.17 & 0.27 \\ 
  LPXBHR24 & 0.00 & 0.03 & 0.04 & 0.07 & 0.07 & 0.02 & 0.01 & 0.03 \\ 
  \hline
   & LPXBHR17 & LPXBHR18 & LPXBHR19 & LPXBHR20 & LPXBHR21 & LPXBHR22 & LPXBHR23 & LPXBHR24\\ 
  \hline
  LPXBHR01 & 0.13 & 0.13 & 0.15 & 0.10 & 0.06 & 0.11 & 0.10 & 0.00 \\ 
  LPXBHR02 & 0.08 & 0.07 & 0.08 & 0.07 & 0.14 & 0.17 & 0.22 & 0.38 \\ 
  LPXBHR03 & 0.09 & 0.08 & 0.07 & 0.06 & 0.11 & 0.12 & 0.15 & 0.42 \\ 
  LPXBHR04 & 0.16 & 0.16 & 0.18 & 0.15 & 0.17 & 0.16 & 0.18 & 0.36 \\ 
  LPXBHR05 & 0.17 & 0.15 & 0.16 & 0.14 & 0.19 & 0.18 & 0.22 & 0.50 \\ 
  LPXBHR06 & 0.22 & 0.20 & 0.21 & 0.19 & 0.25 & 0.24 & 0.27 & 0.55 \\ 
  LPXBHR07 & 0.20 & 0.19 & 0.23 & 0.22 & 0.20 & 0.18 & 0.16 & 0.17 \\ 
  LPXBHR08 & 0.20 & 0.16 & 0.18 & 0.18 & 0.21 & 0.20 & 0.17 & 0.20 \\ 
  LPXBHR09 & 0.24 & 0.17 & 0.20 & 0.19 & 0.17 & 0.18 & 0.10 & 0.00 \\ 
  LPXBHR10 & \textbf{0.67} &  \textbf{0.65} &  \textbf{0.61} &  \textbf{0.63} & 0.55 & 0.50 & 0.39 & 0.04 \\ 
  LPXBHR11 &  \textbf{0.71} &  \textbf{0.65} &  \textbf{0.62} &  \textbf{0.60} & 0.53 & 0.46 & 0.40 & 0.05 \\ 
  LPXBHR12 &  \textbf{0.72} &  \textbf{0.65} &  \textbf{0.61} &  \textbf{0.58} & 0.53 & 0.42 & 0.43 & 0.08 \\ 
  LPXBHR13 &  \textbf{0.78} &  \textbf{0.70} &  \textbf{0.63} &  \textbf{0.52} & 0.44 & 0.35 & 0.39 & 0.08 \\ 
  LPXBHR14 & 0.42 & 0.41 & 0.41 & 0.30 & 0.22 & 0.16 & 0.21 & 0.07 \\ 
  LPXBHR15 & 0.38 & 0.37 & 0.36 & 0.25 & 0.18 & 0.13 & 0.15 & 0.05 \\ 
  LPXBHR16 & 0.46 & 0.46 & 0.42 & 0.36 & 0.28 & 0.26 & 0.20 & 0.02 \\ 
  LPXBHR17 & 1.00 & 0.93 & 0.74 & 0.62 & 0.46 & 0.36 & 0.27 & 0.02 \\ 
  LPXBHR18 & 0.93 & 1.00 & 0.83 & 0.68 & 0.46 & 0.38 & 0.30 & 0.02 \\ 
  LPXBHR19 & 0.74 & 0.83 & 1.00 & 0.81 & 0.59 & 0.45 & 0.46 & 0.05 \\ 
  LPXBHR20 & 0.62 & 0.68 & 0.81 & 1.00 & 0.82 & 0.64 & 0.56 & 0.08 \\ 
  LPXBHR21 & 0.46 & 0.46 & 0.59 & 0.82 & 1.00 & 0.82 & 0.71 & 0.27 \\ 
  LPXBHR22 & 0.36 & 0.38 & 0.45 & 0.64 & 0.82 & 1.00 & 0.81 & 0.23 \\ 
  LPXBHR23 & 0.27 & 0.30 & 0.46 & 0.56 & 0.71 & 0.81 & 1.00 & 0.38 \\ 
  LPXBHR24 & 0.03 & 0.03 & 0.06 & 0.08 & 0.22 & 0.19 & 0.31 & 0.82 \\
\hline\hline
\caption{Population connectedness table for LPXBHR contracts.}
\label{tab: lptab1}
\end{longtable}

\begin{table}[h]
\centering
\scriptsize
\begin{tabular}{r|rrrrrr}
  \hline  \hline
 & LPXBHB & LPXBHOP1 & LPXBHOP2 & LPXBHP & LPXBHRB & LPXBHRP \\ 
  \hline
LPXBHR01 & 0.29 & 0.49 & 0.10 & 0.10 & 0.29 & 0.18 \\ 
LPXBHR02 & 0.30 & 0.48 & 0.12 & 0.14 & 0.30 & 0.18 \\ 
LPXBHR03 & 0.28 & 0.50 & 0.07 & 0.04 & 0.28 & 0.18 \\ 
LPXBHR04 & 0.38 & 0.54 & 0.10 & 0.12 & 0.38 & 0.27 \\ 
LPXBHR05 & 0.53 & 0.78 & 0.15 & 0.14 & 0.53 & 0.40 \\ 
LPXBHR06 & 0.66 & 0.75 & 0.24 & 0.08 & 0.66 & 0.55 \\ 
LPXBHR07 & 0.43 & 0.43 & 0.20 & 0.01 & 0.43 & 0.36 \\ 
LPXBHR08 & 0.44 & 0.40 & 0.20 & 0.01 & 0.44 & 0.39 \\ 
LPXBHR09 & 0.42 & 0.36 & 0.20 & 0.01 & 0.42 & 0.37 \\ 
LPXBHR10 & \textbf{0.91} & 0.67 & 0.49 & 0.29 & \textbf{0.91} & \textbf{0.87} \\ 
LPXBHR11 & \textbf{0.93} & 0.68 & 0.48 & 0.37 & \textbf{0.93} & \textbf{0.90} \\ 
LPXBHR12 & \textbf{0.88} & 0.59 & 0.48 & 0.49 & \textbf{0.88} & \textbf{0.88} \\ 
LPXBHR13 & \textbf{0.84} & 0.54 & 0.42 & 0.48 & \textbf{0.84} & \textbf{0.88} \\ 
LPXBHR14 & 0.49 & 0.29 & 0.22 & 0.33 & 0.49 & 0.56 \\ 
LPXBHR15 & 0.43 & 0.23 & 0.17 & 0.21 & 0.43 & 0.51 \\ 
LPXBHR16 & 0.52 & 0.29 & 0.31 & 0.14 & 0.52 & 0.59 \\ 
LPXBHR17 & 0.79 & 0.48 & 0.35 & 0.24 & 0.79 &\textbf{ 0.89} \\ 
LPXBHR18 & 0.77 & 0.44 & 0.37 & 0.24 & 0.77 & \textbf{0.87} \\ 
LPXBHR19 & 0.75 & 0.43 & 0.50 & 0.30 & 0.75 & \textbf{0.81} \\ 
LPXBHR20 & 0.72 & 0.37 & 0.65 & 0.31 & 0.72 & 0.73 \\ 
LPXBHR21 & 0.65 & 0.33 & \textbf{0.87} & 0.28 & 0.65 & 0.61 \\ 
LPXBHR22 & 0.61 & 0.36 & \textbf{0.91} & 0.16 & 0.61 & 0.50 \\ 
LPXBHR23 & 0.51 & 0.26 & \textbf{0.93} & 0.34 & 0.51 & 0.44 \\ 
LPXBHR24 & 0.07 & 0.11 & 0.41 & 0.10 & 0.07 & 0.06 \\ 
   \hline  \hline
\end{tabular} 
\caption{The ``To'' impacts from the six LPXBHxx-type power contracts to the 24 LPXBHR-type contracts. The numbers larger than $0.8$ are marked in bold font.} \label{tab: to24}
\end{table}

\begin{figure}[h!]
\hspace*{0.7cm}
\centering
\includegraphics[scale=0.7]{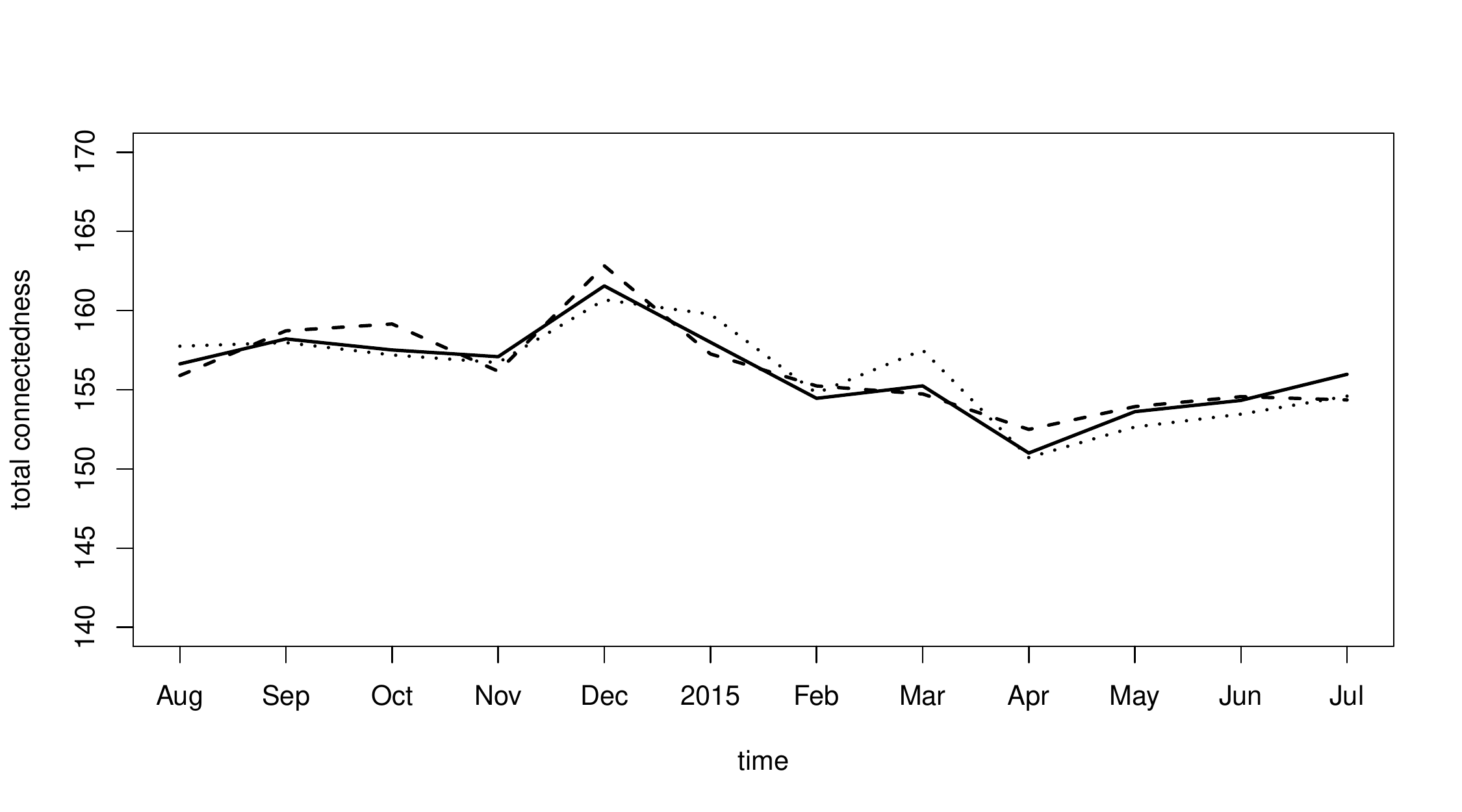}
\caption{The time-varying full sample system-wide connectedness estimated at different forecast horizons, from August 2014 to July 2015. The solid line is $C_{h_{0},t}$, the dashed line is $C_{h_{1},t}$, and  the dotted line is $C_{h_{2},t}$.}
\label{fig: rob1}
\end{figure}

\end{document}